\documentclass[aps,prl,twocolumn,showpacs,superscriptaddress,nofootinbib,floatfix]{revtex4-1}
\usepackage{amsfonts}
\usepackage{tikz}
\usepackage{pgfplots}
\usepackage{amsmath}
\usepackage{amssymb}
\usepackage{bm}
\usepackage{dcolumn}
\usepgfplotslibrary{groupplots}
\usepackage{wrapfig}
\usepackage{graphicx}
\usepackage{graphics}
\usepackage{latexsym}
\usepackage{rotating}
\usepackage[colorlinks=true]{hyperref}
\usepackage[all]{hypcap} 
\usepackage{xspace} 
\usepackage{xcolor}
\usepackage{mathrsfs}
\usepackage{siunitx}
\usepackage{multirow}
\usepackage{soul}
\usepackage{lipsum}
\usepackage{colortbl}
\usepackage{ulem}
\usepgfplotslibrary{fillbetween}
\definecolor {darkgreen}{rgb}{0.2,0.7,0.2}

\newcommand\be{\begin{equation}}
\newcommand\ee{\end{equation}}
\newcommand\bw{\begin{widetext}}
\newcommand\ew{\end{widetext}}

\newcommand{\bea}{\begin{eqnarray}}
\newcommand{\eea}{\end{eqnarray}}

\newcommand{\cm}{{\rm cm}}

\newcommand{\g}{{\rm g}}

\newcommand{\Mo}{{M_{\odot}}}

\usepackage{cellspace} %
\newlength\figureheight
\newlength\figurewidth

\newlength\figureheightmed
\newlength\figurewidthmed

\newlength\figureheightlarge
\newlength\figurewidthlarge

\begin{document}
\title{Neutrino trapping and out-of-equilibrium effects\\
  in binary neutron star merger remnants}

\author{Pedro L. Espino}
\affiliation{Institute for Gravitation \& the Cosmos, The Pennsylvania 
State University, University Park PA 16802, USA}
\affiliation{Department of Physics, University of California, Berkeley, 
CA 94720, USA}

\author{Peter Hammond}
\affiliation{Institute for Gravitation \& the Cosmos, The Pennsylvania 
State University, University Park PA 16802, USA}

\author{David Radice}
\affiliation{Institute for Gravitation \& the Cosmos, The Pennsylvania 
State University, University Park PA 16802, USA}
\affiliation{Department of Physics, The Pennsylvania State University, 
University Park PA 16802, USA}
\affiliation{Department of Astronomy \& Astrophysics, The Pennsylvania 
State University, University Park PA 16802, USA}

\author{Sebastiano \surname{Bernuzzi}}
\affiliation{Theoretisch-Physikalisches Institut, Friedrich-Schiller-Universit{\"a}t Jena, 07743, Jena, Germany}

\author{Rossella \surname{Gamba}}
\affiliation{Theoretisch-Physikalisches Institut, Friedrich-Schiller-Universit{\"a}t Jena, 07743, Jena, Germany}
\author{Francesco \surname{Zappa}}
\affiliation{Theoretisch-Physikalisches Institut, Friedrich-Schiller-Universit{\"a}t Jena, 07743, Jena, Germany}
\author{Lu\'{i}s Felipe \surname{Longo Micchi}}
\affiliation{Theoretisch-Physikalisches Institut, Friedrich-Schiller-Universit{\"a}t Jena, 07743, Jena, Germany}
\author{Albino Perego}
\affiliation{Dipartimento di Fisica, Università di Trento, Via Sommarive 14, 38123 Trento, Italy}
\affiliation{
INFN-TIFPA, Trento Institute for Fundamental Physics and Applications, via Sommarive 14, I-38123 Trento, Italy}

\begin{abstract}
We study out-of-thermodynamic equilibrium effects in neutron star
mergers with 3D general-relativistic neutrino-radiation large-eddy
simulations. During merger, the cores of the neutron stars remain cold
($T  \sim$ a few MeV) and out of thermodynamic equilibrium with trapped
neutrinos originating from the hot collisional interface between the
stars.
However, within ${\sim}2{-}3$ milliseconds matter and neutrinos
reach equilibrium everywhere in the remnant. Our results show that
dissipative effects, such as bulk viscosity, if present, are only active
for a short window of time after the merger.
\end{abstract}

\date{\today} \maketitle

\paragraph*{Introduction.---}
Neutrino interactions are expected to have profound implications for the 
multimessenger signals 
associated with binary neutron star (BNS) mergers. Not only do we expect significant 
neutrino luminosities from BNS mergers~\cite{Eichler:1989ve,Ruffert:1996by, Rosswog:2003rv, 
Sekiguchi_2015, Palenzuela_2015, Foucart_2016,PhysRevD.93.044019, 
PhysRevD.96.123015,PhysRevD.102.103015,Burrows:2019zce,Kullmann:2021gvo,Cusinato:2021zin, Radice:2021jtw} but neutrinos may also have a significant impact on
the properties 
of other BNS merger observables such as kilonovae (KN) and gravitational waves (GWs). 
For example, the conditions following a BNS merger 
allow for short-enough neutrino mean free paths to result in significant neutrino 
re-absorption into the medium. Such re-absorption is expected to result in the 
systematic increase of the average electron fraction of the 
post-merger disk and 
ejecta~\cite{Wanajo:2014wha,Perego:2014fma,PhysRevD.93.044019,Radice:2021jtw,
Zappa:2022rpd}, thereby 
affecting the 
r-process nucleosynthetic 
yields and subsequently the KN properties~\cite{Metzger:2014ila,
Martin:2015hxa,Perego:2017wtu,Radice:2018pdn,Zappa:2022rpd}.
While eventually neutrinos stream away from the system on the diffusion timescale, 
they can be temporarily considered as trapped if the diffusion timescale becomes much 
larger than the dynamical timescale. If production and absorption reactions are fast 
enough, they can equilibrate with matter inside the system.
The timescale
for neutrino interactions may be commensurate with that of 
the local fluid dynamics (i.e., local fluid compression/rarefaction 
driven by 
oscillations of the post-merger remnant massive neutron star 
(RMNS))~\cite{Schmitt:2017efp, Alford:2019kdw, Camelio:2022ljs, Camelio:2022fds}.
%
It has been suggested that the dissipative, out-of-equilibrium effects
in the matter-neutrino mixture produced in mergers may result in
significant bulk-viscous damping of the post-merger
oscillations of the RMNS \cite{Alford:2017rxf, Schmitt:2017efp,
Alford:2019kdw, Endrizzi:2019trv, Hammond:2021vtv, Most:2022yhe,
Camelio:2022ljs, Camelio:2022fds} and, in turn, of the post-merger GW
amplitude~\cite{Most:2022yhe}. 
Microphysical neutrino interactions are a key ingredient in  
BNS mergers and must be considered for a full picture of the relevant astrophysical 
phenomena and, crucially, the re-absorption of neutrinos must be accurately modeled~\cite{Zappa:2022rpd, LongoMicchi:2023khv}.
To this end, moment-based neutrino transport schemes are well-suited to capturing the 
aforementioned phenomena, as they provide an accurate method for capturing the 
effects of neutrino re-absorption in the medium
across all opacity regimes~\cite{Shibata:2011kx, Sekiguchi_2015, 
Foucart_2015, Radice:2016dwd}.

In this work we study the effect of trapped neutrinos in the core of the RMNSs on the 
local chemical equilibrium.
We consider a total of sixteen 3D general relativistic radiation hydrodynamics 
(GRRHD) simulations across 
equation of state (EOS) models, mass ratios, and grid resolutions. 
We use the {\tt THC\_M1} 
code~\cite{Radice:2021jtw}, which 
implements M1 neutrino transport within the well-established {\tt WhiskyTHC} 
code~\cite{WHiskyTHC1, WhiskyTHC2} and 
consider several qualitative and quantitative diagnostics to understand the potential 
size of out-of-equilibrium effects in the post-merger environment. We find that \textit{the 
presence of trapped neutrinos implies that not all neutrinos produced during and 
after the merger efficiently free-stream away from the system. They are instead 
available for 
interactions which may drive the matter toward local weak equilibrium}. 
Although neutrino trapping in BNS mergers with M1 neutrino transport has been 
previously considered~\cite{Foucart:2015gaa}, our study provides the 
first consideration of the potential impact of out-of-equilibrium effects in the 
post-merger environment using ab-initio simulations with M1 neutrino transport and 
depicts a nuanced picture of the effects of neutrino interactions in BNS mergers.
Unless otherwise noted, we assume natural, geometrized units where $G=c=k_{\rm B}=1$.

\paragraph*{Methods and Diagnostics.---}

We consider numerical relativity simulations of 8 binaries with total mass
$M=2.68\Mo$, mass ratio of either $q=1$ or $q=1.2$, and with matter described by 4 microphysical EOS models.
The EOS models in our study include BLh~\cite{Bombaci:2018ksa}, 
DD2~\cite{Hempel2010}, SFHo~\cite{Steiner_2010}, and
SLy4~\cite{Chabanat:1997un,PhysRevC.96.065802}.
The non-rotating, equilibrium neutron stars described by these EOS models are broadly 
compatible with various astrophysical constrains, including those from
massive pulsars~\cite{Cromartie:2019kug,Fonseca:2021wxt},
GW170817~\cite{Abbott:2018exr} and NICER~\cite{Miller:2021qha,Riley:2021pdl}.
Simulations are performed by consistently evolving constraint-satisfying initial data 
in quasi circular orbits through merger. We use the 3+1 Z4c scheme to solve
Einstein's equations \citep{Bernuzzi:2009ex,Hilditch:2012fp} coupled to
the general relativistic hydrodynamics evolution and the truncated,
two-moment M1 gray (energy-integrated) scheme with the Minerbo closure as 
implemented in the {\tt THC\_M1} code~\cite{Radice:2021jtw}.
The \texttt{WhiskyTHC}~code (of which the {\tt THC\_M1} code is an extension)
\citep{Radice:2012cu,Radice:2013hxh,Radice:2013xpa,Radice:2015nva,
Radice:2016dwd,Radice:2021jtw} is built upon the \texttt{Cactus} 
framework \citep{Font:2001ew,Schnetter:2007rb}. 
The simulation domain is a cube of side ${\sim}3024$ km
centered at the center of mass of the binary system. Within this cube, we consider 
three nested cubes (two of which track each binary component, while the other is 
fixed at the computational grid origin) with 7 levels of grid 
refinement such that the finest-level grid resolution is 
${\Delta x}_\mathrm{SR} \approx \SI{185}{m}$; we refer to simulations with this grid 
resolution as standard-resolution (SR). 
To understand the effects of grid resolution, we also consider simulations at
a coarser finest-level grid resolution of 
${\Delta x}_\mathrm{LR} \approx 247$~m, which we refer to as low-resolution (LR). 
Ours is the 
largest-to-date sample of
binaries simulated with the M1 neutrino transport scheme and the first consideration 
of out-of-equilibrium effects using this scheme.
The {\tt THC\_M1} code considers the evolution of three neutrino species including 
the electron neutrino $\nu_e$, electron anti-neutrino $\bar{\nu}_e$, and a collective 
species describing heavy flavor neutrinos 
and antineutrinos $\nu_x$. 
The set of weak reactions modeled in our simulations is described in detail 
in~\cite{Perego:2019adq}.

In order to qualitatively understand the potential impact of out-of-equilibrium 
effects, we consider several post-process diagnostics calculated using the output of 
our dynamical simulations.
For instance, the impact of out-of-equilibrium matter effects in the remnant may be 
captured by the dimensionless parameter 
$\mathcal{A}=\mu_\Delta/T$~\cite{Hammond:2021vtv},
where $T$ is the temperature and $\mu_\Delta$ is the out-of-equilibrium chemical 
potential. In scenarios where neutrinos stream away on a timescale much smaller than 
any other relevant timescale and are not expected to be part of the system, the local 
equilibrium conditions are those of neutrino-less $\beta$-equilibrium. 
As such, the equilibrium value of $\mu_\Delta$ is~\cite{Hammond:2021vtv}
$\mu_\Delta^{npe} \equiv \mu_{n} - \mu_{p} - \mu_{e}$, 
where $\mu_{i}$ is the chemical potential for species $i$, and the labels 
$n$, $p$, and $e$ correspond to the neutron, proton, and electron, 
respectively. 
In our work, we consider a more accurate measure of
the deviation from local weak equilibrium by including trapped
electron anti-neutrinos. Specifically, in cases where neutrinos may be significantly 
trapped, the local equilibrium condition is that of full $\beta$-equilibrium, and as 
such $\mu_\Delta$ becomes
$\mu_\Delta^{npe\nu} \equiv \mu_{n} - \mu_{p} - \mu_{e} - \mu_{\bar{\nu}_{e}}$.
The chemical potential for the 
proton, neutron, and electron are provided by the EOS models
we employ for each simulation. However, these EOS models do not
explicitly account for the impact of neutrinos, and as such we must
estimate $\mu_\nu$ using the local neutrino fraction $Y_\nu$ extracted from our simulations~\cite{LongoMicchi:2023khv}.
We note that for the analyses presented in this work we typically consider the 
electron anti-neutrino fraction, but we find consistent results when considering the 
electron neutrino fraction with appropriate sign changes instead. We emphasize that 
our M1 scheme is able to self-consistently evolve the neutrino 
fraction $Y_\nu$ and energy $u_\nu$, 
and that electron anti-neutrinos are the most abundant 
species~\cite{Foucart:2015gaa, Perego:2019adq, Zappa:2022rpd} with typical fractions 
of ${\sim}10^{-3}$ in the relevant high-density regions.

\begin{figure*}[ht]
\includegraphics[scale=0.315]{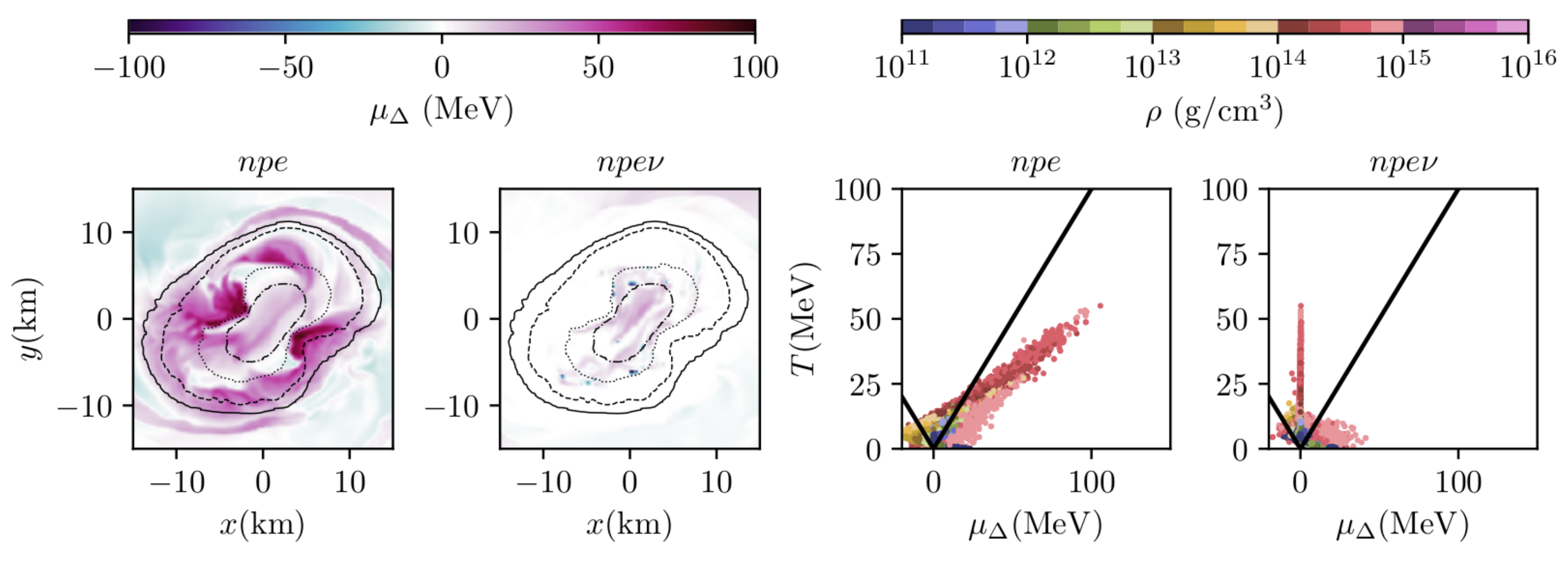}
\centering
\caption{
{\it Left:} Equatorial snapshots of the out-of-equilibrium chemical potential 
$\mu_\Delta$ at a time after the merger of $\delta t\approx \SI{3.69}{ms}$ 
for the equal mass, SR simulation 
in our work with use of the DD2 EOS. 
The leftmost and second-from-left panels depict the case where we assume 
the matter to consist of $npe$ and $npe\nu$ matter, respectively.
We highlight regions in 
the post-merger environment with rest mass density 
$\rho = 0.5\rho_{\rm sat}$, $\rho_{\rm sat}$, 
$2\rho_{\rm sat}$, and $2.5\rho_{\rm sat}$ 
(where $\rho_{\rm sat} \approx \SI{2.5e14}{\g\per\cm\cubed}$ is the nuclear 
saturation density) using solid, dashed, dotted, and dash-
dotted black lines, respectively. {\it Right:} Phase-space histograms for the two 
simulations depicted in the left panel. For these histograms we focus on a time 
window of $\sim \SI{3}{ms}$ before and after the merger. We mark the condition where 
$\mu_\Delta = T$ with solid black lines. Phase-space regions above 
(below) the solid lines where $T=\mu_\Delta$ imply that the matter is close to 
(significantly deviating from) weak equilibrium.
}
\label{fig:betaequil_contours}
\end{figure*}


For a complementary look at the potential impact of out-of-equilibrium effects, we 
also consider the ratio of timescales relevant to bulk viscosity. 
Specifically, bulk viscosity is expected to emerge as a resonant phenomenon when the 
timescales of the local compression and rarefaction of the fluid is commensurate with 
the timescale on which microphysical reactions allows the fluid to reach chemical 
equilibrium~\cite{Schmitt:2017efp, Alford:2019kdw, Camelio:2022ljs, Camelio:2022fds}. 
In the case of the environment following a BNS merger, the relevant timescales 
are those of local density oscillations $\tau_{\rm h}$ and weak interactions 
$\tau_{\nu}$. If these timescales are very disparate, the fluid may undergo adiabatic 
changes in state variables 
either so fast that the composition 
remains effectively fixed ($\tau_\nu \gg \tau_h$), or so slowly that it 
is at all times in chemical equilibrium ($\tau_\nu \ll \tau_h$)
However, when the two timescales are similar the system may 
undergo local density oscillations while similarly undergoing local changes in the 
fluid chemical composition (and thereby local pressure), leading to an emergent 
dissipation~\cite{Camelio:2022ljs, Camelio:2022fds}.

\paragraph*{Results.---}
\begin{figure*}[ht]
\includegraphics[scale=1.05]{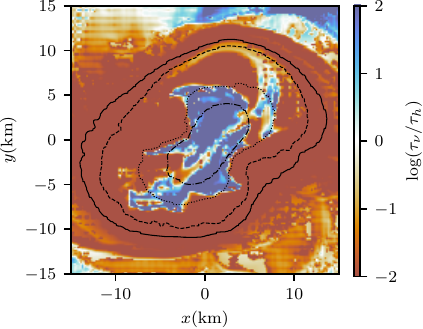}
\includegraphics[scale=0.75]{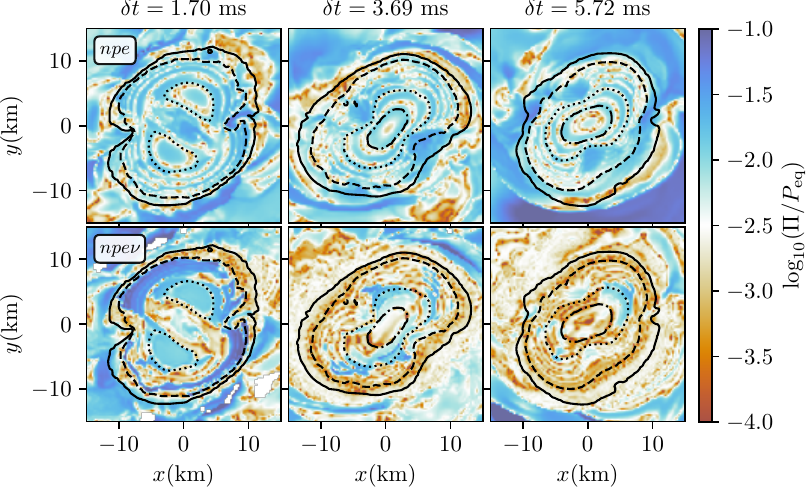}
\centering
\caption{
{\it Left:} Equatorial snapshot of the ratio of relevant timescales at a time after 
merger of $\delta t= \SI{3.69}{ms}$  for the same simulation depicted in 
Fig.~\ref{fig:betaequil_contours}. Specifically, we show the ratio of 
the timescale associated with neutrino absorption to that associated with density 
oscillations.
{\it Right:} Equatorial snapshots of the relative difference between the simulation 
pressure and the equilibrium pressure $\Pi \equiv \lvert P - P_{\rm eq} \rvert$ for 
the same simulation depicted in the left panel. The top panels shows cases where the 
equilibrium pressure is calculated assuming $npe$ matter (i.e., using the the local 
electron fraction corresponding to \textit{neutrino-less} $\beta$-equilibrium).
The bottom panels shows cases where the equilibrium pressure is calculated assuming 
$npe\nu$ matter, using Eqs.~\eqref{eq:u_total_def}-\eqref{eq:Ylep_def}.
}
\label{fig:pressure_eq}
\end{figure*}
In Fig.~\ref{fig:betaequil_contours} we show the 
out-of-equilibrium chemical potential in the equatorial plane of the RMNS for the 
equal mass ratio, SR simulation with use of the DD2 EOS, at a representative time 
from merger of 
$\delta t \approx \SI{3.76}{ms}$ post-merger.
The figure shows that 
matter is not in neutrino-less $\beta$-equilibrium, in line 
with the findings of previous works~\cite{Hammond:2021vtv, Endrizzi:2019trv}.
As we account for neutrinos in our simulation, we can accurately asses whether or not 
matter is in neutrino-trapped equilibrium. We find indications of 
significant neutrino trapping throughout the 
RMNS, with typical neutrino fractions of $Y_\nu \approx 10^{-3}$. 
As such, an
assumption of $npe$ matter is not accurate and regions with  
high opacity remain near local weak 
equilibrium~\cite{Perego:2019adq,Radice:2021jtw,Zappa:2022rpd}.
For 
the sake of brevity, we showcase the results for a single SR simulation, but we find 
all of our findings to hold regardless of EOS model, mass ratio, and grid 
resolution.

In Fig.~\ref{fig:betaequil_contours} we also show the 
phase-space histograms, specifically the $\mu_\Delta$-$T$ plane, for each of the  
cases depicted in the left panels.  
The rightmost panel of 
Fig.~\ref{fig:betaequil_contours} corroborates our finding that when neutrino 
trapping is accounted for, the fluid is driven toward 
$\beta$-equilibrium on timescales that are shorter than the dynamical 
time~\cite{Alford:2019kdw}, thereby potentially reducing the importance of out-of-
equilibrium effects. 
In Fig.~\ref{fig:pressure_eq} we 
show the ratio of neutrino interaction and density oscillation timescales for the 
same simulation and timestamp depicted in Fig.~\ref{fig:betaequil_contours}. 
We estimate the timescale of local density oscillations as 
$\tau_{\rm h} = -\nabla_\mu u^\mu \approx D/\dot{D}$, where $u^\mu$ is the
fluid four-velocity, $D\equiv \rho W \sqrt{\gamma}$, and $\rho$, $W$ and $\gamma$ 
are the rest mass density, Lorentz factor and determinant of the the three-metric, 
respectively.
We estimate the neutrino 
interaction timescale by considering the local scattering $\kappa_{\rm scat}$ and 
absorption $\kappa_{\rm abs}$ opacities, 
$\tau_\nu \approx 1/\sqrt{\kappa_{\rm abs} ( \kappa_{\rm abs} + \kappa_{\rm scat})}.$
Fig.~\ref{fig:pressure_eq} shows
that for the majority of the RMNS 
the neutrino interaction timescale is much 
faster (by at least two orders of magnitude) than the timescale associated with 
density oscillations. In these regions, neutrino interactions happen rapidly enough 
to drive the matter toward local weak equilibrium.
Figs.~\ref{fig:betaequil_contours}-\ref{fig:pressure_eq} show regions with 
significant deviations from weak equilibrium and where the neutrino interaction time 
is significantly smaller than that associated with density oscillations. These are 
likely regions that are in the non-trivial \textit{translucent} regime, which we 
discuss further below.

For a complementary, quantitative consideration of the deviation from local weak 
equilibrium, in Fig.~\ref{fig:pressure_eq} we also show the relative difference 
between the simulation pressure and the pressure corresponding to weak equilibrium 
$\Pi \equiv P(\rho, T, Y_{\rm e}) - P_{\rm eq}(\rho, T^*, Y_{\rm e}^*)$,
where $T^*$ and $Y_{\rm e}^*$ are the inferred 
temperature and electron fraction assuming \textit{neutrino-trapped} local weak 
equilibrium has been established. 
We infer the local values of $T^*$ and $Y_{\rm e}^*$ by 
constructing the neutrino-trapped equilibrium condition in
the microcanonical ensemble, which is relevant since matter in BNS
mergers is not in a thermal bath but fluid elements have evolutions
close to adiabatic. 
Specifically, we assume that 
the total energy density $u_{\rm sim}$ and lepton fraction $Y_{\rm lep, sim}$ 
extracted from our simulations have contributions from trapped neutrinos~\cite{Perego:2019adq} such that
\begin{equation}\label{eq:u_total_def}
u_{\rm sim} = u_{\rm fluid}(\rho, T^*, Y_{\rm e}^*) + u_\nu(\rho, T^*, Y_{\rm e}^*),
\end{equation}
and
\begin{equation}\label{eq:Ylep_def}
Y_{\rm lep, sim} = Y_{\rm e}^* + Y_{\nu_e}(\rho, T^*, Y_{\rm e}^*) - Y_{\bar{\nu}_e}(\rho, T^*, Y_{\rm e}^*),
\end{equation}
where $u_{\rm fluid}$ is the energy density of the fluid as interpolated from the EOS 
model, $u_\nu$ is the contribution to the total energy density from trapped 
neutrinos, and $Y_{\nu_e}$ and $Y_{\bar{\nu}_e}$ are the electron neutrino and 
anti-neutrino fractions, respectively. 
A root-finding algorithm is then used to solve for $T^*$ and $Y_{\rm e}^*$
\footnote{For additional detail on how $Y_{\rm e}^*$, $T^*$, $u_\nu$, 
$Y_{\nu_{\rm e}}$ and $Y_{\bar{\nu}_{\rm e}}$ are calculated, we refer the reader 
to the \textit{Supplemental Material}}.
We do not include the contribution of neutrinos directly on the system 
pressure, as it is negligible. Instead, the impact of neutrinos emerges as a change 
on the local relevant matter conditions (namely $T^*$ and $Y_{\rm e}^*$) 
which in turn modifies the fluid pressure to a new equilibrium state.
In the case in which we treat the equilibrium condition to be that of $npe$ matter,
we find typical pressure deviations from equilibrium of ${\sim} 5\%$ throughout 
all stages of the post-merger evolution. On the other hand, in 
cases where we treat the equilibrium state as including trapped neutrinos we 
generally find smaller deviations from local weak equilibrium, with pressure 
deviations decreasing by at least an order of magnitude in most regions on a 
timescale of approximately $\delta t\SI{3}{ms}$ after the merger.

As suggested by 
Figs.~\ref{fig:betaequil_contours}-\ref{fig:pressure_eq}, there may arise
high-density regions in the RMNS that show significant deviations from local 
neutrino-trapped weak
equilibrium, with $\mu_\Delta \sim \SI{10}{MeV}$, 
$\tau_\nu \gtrsim 10^{2}\tau_{\rm h}$, 
and $\Pi/P_{\rm eq} \sim 0.05$. 
These are regions where matter is out of equilibrium with 
neutrinos and 
 neutrinos are also decoupled. We note that in the translucent 
region neutrinos are trapped due to scattering, so the neutrino-trapped equilibrium 
is the relevant condition.
It is in these translucent
regions where we expect bulk 
viscosity may arise, as neutrino interactions there are not rapid enough to drive the 
matter toward equilibrium faster than the local density oscillations (see the 
{\it Supplemental Material} for additional diagnostics which support this). 
However, these regions are spatially small and typically 
transient, 
existing only for a few ms after the merger. The majority of the RMNS remains 
within 
$\mu_\Delta \lesssim \SI{2}{MeV}$, with 
$\tau_{\nu} \lesssim 10^{-2}\tau_{\rm h}$, and with $\Pi/P_{\rm eq} \sim 10^{-3}$  
during the most violent stages of the 
merger, and the entire RMNS evolves toward the condition of neutrino-trapped weak 
equilibrium on dynamical timescales.
\paragraph*{Conclusion.---}\label{sec:conclusion}

In this letter we have shown, for the first time using M1 neutrino transport in GRRHD 
simulations, that neutrinos may remain significantly trapped in the post-merger 
environment to partake in weak interactions which drive the matter toward weak 
equilibrium. Our simulations indicate that: (1) if 
we account for the presence of trapped neutrinos by considering the neutrino fraction 
$Y_{\nu}$ extracted from our simulations, then the majority of the post-merger system 
remains within $\mu_\Delta \lesssim \SI{2}{MeV}$, with only spatially small regions 
of the remnant \textit{transiently} exceeding these conditions and the RMNS evolving 
toward $\mu_\Delta = 0$ on dynamical timescales; (2) the neutrino interaction 
timescale $\tau_\nu$ is at least two orders of magnitude faster than the timescale 
for local density oscillations $\tau_{\rm h}$ throughout most of the RMNS, with the 
aforementioned regions that show significant deviations from $\mu_\Delta=0$ 
coinciding with regions that have $\tau_\nu \geq \tau_{\rm h}$; 
(3) when considering the matter to 
consist in part of a trapped neutrino gas, the local deviations from the equilibrium 
pressure do not exceed ${\sim} 5\%$ (pressure deviations remain close to ${\sim} 1\%$ for 
most of the RMNS) and these deviations decrease toward 
${\sim} 0.1 \%$ on dynamical timescales as the RMNS continues to evolve.
These analyses show that out-of-equilibrium effects are
likely to be small for most of the post-merger evolution, with a possible 
exception in the translucent region where neutrinos are present but not in thermal or 
chemical equilibrium with the matter. Our simulation draw a
picture of out-of-equilibrium effects in BNS mergers that is much more nuanced than 
previously anticipated.

The key element in our study is the use of an M1 neutrino transport
scheme that does not assume thermodynamic equilibrium between matter and 
radiation. This allows our simulations to capture the trapping of neutrinos and a 
more accurate estimate of the chemical potential, relevant timescales, and deviations 
of the pressure from equilibrium.
Our simulations suggest that the 
post-merger system consists in part of a 
trapped neutrino gas that keeps the matter near local weak equilibrium. A relatively 
cheap and effective way of treating this would be to include trapped neutrinos as 
part of the fluid equation of state.
There are potentially other effects that could enhance 
the bulk viscosity such as the production of thermal 
pions~\cite{Hammond:2021vtv} or muons~\cite{Loffredo:2022prq}, the presence of hyperons~\cite{Alford:2020pld}, or high-density deconfinement phase transitions~\cite{Alford:2013pma, AlfordCSS13}, which 
we do not account for. A consideration of the effects of bulk viscosity 
would require the modeling of microphysics effects including 
additional reactions and degrees of freedom. We leave this to future work.
\paragraph*{Acknowledgments.---}
PE acknowledges funding from the National Science Foundation under Grant
No. PHY-2020275.
PH acknowledges funding from the National Science Foundation under Grant No.~PHY-2116686.
DR acknowledges funding from the U.S. Department of Energy, Office of
Science, Division of Nuclear Physics under Award Number(s) DE-SC0021177,
DE-SC0024388, and from the National Science Foundation under Grants No.
PHY-2011725, PHY-2116686, and AST-2108467.
RG is supported by the Deutsche Forschungsgemeinschaft (DFG) under Grant No.
406116891 within the Research Training Group RTG 2522/1.
FZ acknowledges support from the EU H2020 under ERC Starting
Grant, no.~BinGraSp-714626.  
LFLM acknowledges funding from the EU Horizon under ERC Consolidator Grant, no. InspiReM-101043372
LL SB acknowledges support from the EU Horizon under ERC Consolidator Grant,
no. InspiReM-101043372.
SB acknowledges support from the EU Horizon under ERC Consolidator Grant,
no. InspiReM-101043372 and from the Deutsche Forschungsgemeinschaft
(DFG) project MEMI number BE 6301/2-1.
Simulations were performed on Bridges2, Expanse (NSF XSEDE allocation
TG-PHY160025), Frontera (NSF LRAC allocation PHY23001), and Perlmutter. 
This research used resources of the National Energy Research Scientific
Computing Center, a DOE Office of Science User Facility supported by the
Office of Science of the U.S.~Department of Energy under Contract
No.~DE-AC02-05CH11231.
The authors acknowledge the Gauss Centre for Supercomputing
e.V. (\url{www.gauss-centre.eu}) for funding this project by providing
computing time on the GCS Supercomputer SuperMUC-NG at LRZ
(allocation {\tt pn36ge} and {\tt pn36jo}).

\bibliography{ref}

\def\prd{Phys. Rev. D}\def\prl{Phys. Rev. Lett.}\def\apjl{Astrophys. J.
  Lett.}\def\apjs{Astrophys. J. Suppl.}\def\apj{Astrophys. J.}\def\aj{Astron.
  J.}\def\aap{Astron. Astrophys.}\def\aaps{Astron. Astrophys.
  Suppl.}\def\araa{Ann. Rev. of Astron. and Astrophys.}\def\adp{Ann.
  Phy.}\def\cqg{Classical Quant. Grav.}\def\mnras{Mon. Not. R. Astron.
  Soc.}\def\physrep{Phys. Rep.}\def\nat{Nat.}\def\pasj{Pub. Astron. Soc. of
  Jap.}\def\prc{Phys. Rev. C}\def\sovast{Soviet. Ast.}
\begin{thebibliography}{58}%
\makeatletter
\providecommand \@ifxundefined [1]{%
 \@ifx{#1\undefined}
}%
\providecommand \@ifnum [1]{%
 \ifnum #1\expandafter \@firstoftwo
 \else \expandafter \@secondoftwo
 \fi
}%
\providecommand \@ifx [1]{%
 \ifx #1\expandafter \@firstoftwo
 \else \expandafter \@secondoftwo
 \fi
}%
\providecommand \natexlab [1]{#1}%
\providecommand \enquote  [1]{``#1''}%
\providecommand \bibnamefont  [1]{#1}%
\providecommand \bibfnamefont [1]{#1}%
\providecommand \citenamefont [1]{#1}%
\providecommand \href@noop [0]{\@secondoftwo}%
\providecommand \href [0]{\begingroup \@sanitize@url \@href}%
\providecommand \@href[1]{\@@startlink{#1}\@@href}%
\providecommand \@@href[1]{\endgroup#1\@@endlink}%
\providecommand \@sanitize@url [0]{\catcode `\\12\catcode `\$12\catcode
  `\&12\catcode `\#12\catcode `\^12\catcode `\_12\catcode `\%12\relax}%
\providecommand \@@startlink[1]{}%
\providecommand \@@endlink[0]{}%
\providecommand \url  [0]{\begingroup\@sanitize@url \@url }%
\providecommand \@url [1]{\endgroup\@href {#1}{\urlprefix }}%
\providecommand \urlprefix  [0]{URL }%
\providecommand \Eprint [0]{\href }%
\providecommand \doibase [0]{http://dx.doi.org/}%
\providecommand \selectlanguage [0]{\@gobble}%
\providecommand \bibinfo  [0]{\@secondoftwo}%
\providecommand \bibfield  [0]{\@secondoftwo}%
\providecommand \translation [1]{[#1]}%
\providecommand \BibitemOpen [0]{}%
\providecommand \bibitemStop [0]{}%
\providecommand \bibitemNoStop [0]{.\EOS\space}%
\providecommand \EOS [0]{\spacefactor3000\relax}%
\providecommand \BibitemShut  [1]{\csname bibitem#1\endcsname}%
\let\auto@bib@innerbib\@empty
\bibitem [{\citenamefont {Eichler}\ \emph {et~al.}(1989)\citenamefont
  {Eichler}, \citenamefont {Livio}, \citenamefont {Piran},\ and\ \citenamefont
  {Schramm}}]{Eichler:1989ve}%
  \BibitemOpen
  \bibfield  {author} {\bibinfo {author} {\bibfnamefont {D.}~\bibnamefont
  {Eichler}}, \bibinfo {author} {\bibfnamefont {M.}~\bibnamefont {Livio}},
  \bibinfo {author} {\bibfnamefont {T.}~\bibnamefont {Piran}}, \ and\ \bibinfo
  {author} {\bibfnamefont {D.~N.}\ \bibnamefont {Schramm}},\ }\href {\doibase
  10.1038/340126a0} {\bibfield  {journal} {\bibinfo  {journal} {Nature}\
  }\textbf {\bibinfo {volume} {340}},\ \bibinfo {pages} {126} (\bibinfo {year}
  {1989})}\BibitemShut {NoStop}%
\bibitem [{\citenamefont {Ruffert}\ \emph {et~al.}(1997)\citenamefont
  {Ruffert}, \citenamefont {Janka}, \citenamefont {Takahashi},\ and\
  \citenamefont {Schaefer}}]{Ruffert:1996by}%
  \BibitemOpen
  \bibfield  {author} {\bibinfo {author} {\bibfnamefont {M.}~\bibnamefont
  {Ruffert}}, \bibinfo {author} {\bibfnamefont {H.~T.}\ \bibnamefont {Janka}},
  \bibinfo {author} {\bibfnamefont {K.}~\bibnamefont {Takahashi}}, \ and\
  \bibinfo {author} {\bibfnamefont {G.}~\bibnamefont {Schaefer}},\ }\href@noop
  {} {\bibfield  {journal} {\bibinfo  {journal} {Astron. Astrophys.}\ }\textbf
  {\bibinfo {volume} {319}},\ \bibinfo {pages} {122} (\bibinfo {year}
  {1997})},\ \Eprint {http://arxiv.org/abs/astro-ph/9606181}
  {arXiv:astro-ph/9606181} \BibitemShut {NoStop}%
\bibitem [{\citenamefont {Rosswog}\ and\ \citenamefont
  {Liebendoerfer}(2003)}]{Rosswog:2003rv}%
  \BibitemOpen
  \bibfield  {author} {\bibinfo {author} {\bibfnamefont {S.}~\bibnamefont
  {Rosswog}}\ and\ \bibinfo {author} {\bibfnamefont {M.}~\bibnamefont
  {Liebendoerfer}},\ }\href {\doibase 10.1046/j.1365-8711.2003.06579.x}
  {\bibfield  {journal} {\bibinfo  {journal} {Mon. Not. Roy. Astron. Soc.}\
  }\textbf {\bibinfo {volume} {342}},\ \bibinfo {pages} {673} (\bibinfo {year}
  {2003})},\ \Eprint {http://arxiv.org/abs/astro-ph/0302301}
  {arXiv:astro-ph/0302301} \BibitemShut {NoStop}%
\bibitem [{\citenamefont {Sekiguchi}\ \emph {et~al.}(2015)\citenamefont
  {Sekiguchi}, \citenamefont {Kiuchi}, \citenamefont {Kyutoku},\ and\
  \citenamefont {Shibata}}]{Sekiguchi_2015}%
  \BibitemOpen
  \bibfield  {author} {\bibinfo {author} {\bibfnamefont {Y.}~\bibnamefont
  {Sekiguchi}}, \bibinfo {author} {\bibfnamefont {K.}~\bibnamefont {Kiuchi}},
  \bibinfo {author} {\bibfnamefont {K.}~\bibnamefont {Kyutoku}}, \ and\
  \bibinfo {author} {\bibfnamefont {M.}~\bibnamefont {Shibata}},\ }\href
  {\doibase 10.1103/physrevd.91.064059} {\bibfield  {journal} {\bibinfo
  {journal} {Physical Review D}\ }\textbf {\bibinfo {volume} {91}} (\bibinfo
  {year} {2015}),\ 10.1103/physrevd.91.064059}\BibitemShut {NoStop}%
\bibitem [{\citenamefont {Palenzuela}\ \emph {et~al.}(2015)\citenamefont
  {Palenzuela}, \citenamefont {Liebling}, \citenamefont {Neilsen},
  \citenamefont {Lehner}, \citenamefont {Caballero}, \citenamefont {O?Connor},\
  and\ \citenamefont {Anderson}}]{Palenzuela_2015}%
  \BibitemOpen
  \bibfield  {author} {\bibinfo {author} {\bibfnamefont {C.}~\bibnamefont
  {Palenzuela}}, \bibinfo {author} {\bibfnamefont {S.~L.}\ \bibnamefont
  {Liebling}}, \bibinfo {author} {\bibfnamefont {D.}~\bibnamefont {Neilsen}},
  \bibinfo {author} {\bibfnamefont {L.}~\bibnamefont {Lehner}}, \bibinfo
  {author} {\bibfnamefont {O.}~\bibnamefont {Caballero}}, \bibinfo {author}
  {\bibfnamefont {E.}~\bibnamefont {O?Connor}}, \ and\ \bibinfo {author}
  {\bibfnamefont {M.}~\bibnamefont {Anderson}},\ }\href {\doibase
  10.1103/physrevd.92.044045} {\bibfield  {journal} {\bibinfo  {journal}
  {Physical Review D}\ }\textbf {\bibinfo {volume} {92}} (\bibinfo {year}
  {2015}),\ 10.1103/physrevd.92.044045}\BibitemShut {NoStop}%
\bibitem [{\citenamefont {Foucart}\ \emph
  {et~al.}(2016{\natexlab{a}})\citenamefont {Foucart}, \citenamefont
  {O?Connor}, \citenamefont {Roberts}, \citenamefont {Kidder}, \citenamefont
  {Pfeiffer},\ and\ \citenamefont {Scheel}}]{Foucart_2016}%
  \BibitemOpen
  \bibfield  {author} {\bibinfo {author} {\bibfnamefont {F.}~\bibnamefont
  {Foucart}}, \bibinfo {author} {\bibfnamefont {E.}~\bibnamefont {O?Connor}},
  \bibinfo {author} {\bibfnamefont {L.}~\bibnamefont {Roberts}}, \bibinfo
  {author} {\bibfnamefont {L.~E.}\ \bibnamefont {Kidder}}, \bibinfo {author}
  {\bibfnamefont {H.~P.}\ \bibnamefont {Pfeiffer}}, \ and\ \bibinfo {author}
  {\bibfnamefont {M.~A.}\ \bibnamefont {Scheel}},\ }\href {\doibase
  10.1103/physrevd.94.123016} {\bibfield  {journal} {\bibinfo  {journal}
  {Physical Review D}\ }\textbf {\bibinfo {volume} {94}} (\bibinfo {year}
  {2016}{\natexlab{a}}),\ 10.1103/physrevd.94.123016}\BibitemShut {NoStop}%
\bibitem [{\citenamefont {Foucart}\ \emph
  {et~al.}(2016{\natexlab{b}})\citenamefont {Foucart}, \citenamefont {Haas},
  \citenamefont {Duez}, \citenamefont {O'Connor}, \citenamefont {Ott},
  \citenamefont {Roberts}, \citenamefont {Kidder}, \citenamefont {Lippuner},
  \citenamefont {Pfeiffer},\ and\ \citenamefont {Scheel}}]{PhysRevD.93.044019}%
  \BibitemOpen
  \bibfield  {author} {\bibinfo {author} {\bibfnamefont {F.}~\bibnamefont
  {Foucart}}, \bibinfo {author} {\bibfnamefont {R.}~\bibnamefont {Haas}},
  \bibinfo {author} {\bibfnamefont {M.~D.}\ \bibnamefont {Duez}}, \bibinfo
  {author} {\bibfnamefont {E.}~\bibnamefont {O'Connor}}, \bibinfo {author}
  {\bibfnamefont {C.~D.}\ \bibnamefont {Ott}}, \bibinfo {author} {\bibfnamefont
  {L.}~\bibnamefont {Roberts}}, \bibinfo {author} {\bibfnamefont {L.~E.}\
  \bibnamefont {Kidder}}, \bibinfo {author} {\bibfnamefont {J.}~\bibnamefont
  {Lippuner}}, \bibinfo {author} {\bibfnamefont {H.~P.}\ \bibnamefont
  {Pfeiffer}}, \ and\ \bibinfo {author} {\bibfnamefont {M.~A.}\ \bibnamefont
  {Scheel}},\ }\href {\doibase 10.1103/PhysRevD.93.044019} {\bibfield
  {journal} {\bibinfo  {journal} {Phys. Rev. D}\ }\textbf {\bibinfo {volume}
  {93}},\ \bibinfo {pages} {044019} (\bibinfo {year}
  {2016}{\natexlab{b}})}\BibitemShut {NoStop}%
\bibitem [{\citenamefont {Wu}\ \emph {et~al.}(2017)\citenamefont {Wu},
  \citenamefont {Tamborra}, \citenamefont {Just},\ and\ \citenamefont
  {Janka}}]{PhysRevD.96.123015}%
  \BibitemOpen
  \bibfield  {author} {\bibinfo {author} {\bibfnamefont {M.-R.}\ \bibnamefont
  {Wu}}, \bibinfo {author} {\bibfnamefont {I.}~\bibnamefont {Tamborra}},
  \bibinfo {author} {\bibfnamefont {O.}~\bibnamefont {Just}}, \ and\ \bibinfo
  {author} {\bibfnamefont {H.-T.}\ \bibnamefont {Janka}},\ }\href {\doibase
  10.1103/PhysRevD.96.123015} {\bibfield  {journal} {\bibinfo  {journal} {Phys.
  Rev. D}\ }\textbf {\bibinfo {volume} {96}},\ \bibinfo {pages} {123015}
  (\bibinfo {year} {2017})}\BibitemShut {NoStop}%
\bibitem [{\citenamefont {George}\ \emph {et~al.}(2020)\citenamefont {George},
  \citenamefont {Wu}, \citenamefont {Tamborra}, \citenamefont
  {Ardevol-Pulpillo},\ and\ \citenamefont {Janka}}]{PhysRevD.102.103015}%
  \BibitemOpen
  \bibfield  {author} {\bibinfo {author} {\bibfnamefont {M.}~\bibnamefont
  {George}}, \bibinfo {author} {\bibfnamefont {M.-R.}\ \bibnamefont {Wu}},
  \bibinfo {author} {\bibfnamefont {I.}~\bibnamefont {Tamborra}}, \bibinfo
  {author} {\bibfnamefont {R.}~\bibnamefont {Ardevol-Pulpillo}}, \ and\
  \bibinfo {author} {\bibfnamefont {H.-T.}\ \bibnamefont {Janka}},\ }\href
  {\doibase 10.1103/PhysRevD.102.103015} {\bibfield  {journal} {\bibinfo
  {journal} {Phys. Rev. D}\ }\textbf {\bibinfo {volume} {102}},\ \bibinfo
  {pages} {103015} (\bibinfo {year} {2020})}\BibitemShut {NoStop}%
\bibitem [{\citenamefont {Burrows}\ \emph {et~al.}(2020)\citenamefont
  {Burrows}, \citenamefont {Radice}, \citenamefont {Vartanyan}, \citenamefont
  {Nagakura}, \citenamefont {Skinner},\ and\ \citenamefont
  {Dolence}}]{Burrows:2019zce}%
  \BibitemOpen
  \bibfield  {author} {\bibinfo {author} {\bibfnamefont {A.}~\bibnamefont
  {Burrows}}, \bibinfo {author} {\bibfnamefont {D.}~\bibnamefont {Radice}},
  \bibinfo {author} {\bibfnamefont {D.}~\bibnamefont {Vartanyan}}, \bibinfo
  {author} {\bibfnamefont {H.}~\bibnamefont {Nagakura}}, \bibinfo {author}
  {\bibfnamefont {M.~A.}\ \bibnamefont {Skinner}}, \ and\ \bibinfo {author}
  {\bibfnamefont {J.}~\bibnamefont {Dolence}},\ }\href {\doibase
  10.1093/mnras/stz3223} {\bibfield  {journal} {\bibinfo  {journal} {Mon. Not.
  Roy. Astron. Soc.}\ }\textbf {\bibinfo {volume} {491}},\ \bibinfo {pages}
  {2715} (\bibinfo {year} {2020})},\ \Eprint {http://arxiv.org/abs/1909.04152}
  {arXiv:1909.04152 [astro-ph.HE]} \BibitemShut {NoStop}%
\bibitem [{\citenamefont {Kullmann}\ \emph {et~al.}(2022)\citenamefont
  {Kullmann}, \citenamefont {Goriely}, \citenamefont {Just}, \citenamefont
  {Ardevol-Pulpillo}, \citenamefont {Bauswein},\ and\ \citenamefont
  {Janka}}]{Kullmann:2021gvo}%
  \BibitemOpen
  \bibfield  {author} {\bibinfo {author} {\bibfnamefont {I.}~\bibnamefont
  {Kullmann}}, \bibinfo {author} {\bibfnamefont {S.}~\bibnamefont {Goriely}},
  \bibinfo {author} {\bibfnamefont {O.}~\bibnamefont {Just}}, \bibinfo {author}
  {\bibfnamefont {R.}~\bibnamefont {Ardevol-Pulpillo}}, \bibinfo {author}
  {\bibfnamefont {A.}~\bibnamefont {Bauswein}}, \ and\ \bibinfo {author}
  {\bibfnamefont {H.~T.}\ \bibnamefont {Janka}},\ }\href {\doibase
  10.1093/mnras/stab3393} {\bibfield  {journal} {\bibinfo  {journal} {Mon. Not.
  Roy. Astron. Soc.}\ }\textbf {\bibinfo {volume} {510}},\ \bibinfo {pages}
  {2804} (\bibinfo {year} {2022})},\ \Eprint {http://arxiv.org/abs/2109.02509}
  {arXiv:2109.02509 [astro-ph.HE]} \BibitemShut {NoStop}%
\bibitem [{\citenamefont {Cusinato}\ \emph {et~al.}(2021)\citenamefont
  {Cusinato}, \citenamefont {Guercilena}, \citenamefont {Perego}, \citenamefont
  {Logoteta}, \citenamefont {Radice}, \citenamefont {Bernuzzi},\ and\
  \citenamefont {Ansoldi}}]{Cusinato:2021zin}%
  \BibitemOpen
  \bibfield  {author} {\bibinfo {author} {\bibfnamefont {M.}~\bibnamefont
  {Cusinato}}, \bibinfo {author} {\bibfnamefont {F.~M.}\ \bibnamefont
  {Guercilena}}, \bibinfo {author} {\bibfnamefont {A.}~\bibnamefont {Perego}},
  \bibinfo {author} {\bibfnamefont {D.}~\bibnamefont {Logoteta}}, \bibinfo
  {author} {\bibfnamefont {D.}~\bibnamefont {Radice}}, \bibinfo {author}
  {\bibfnamefont {S.}~\bibnamefont {Bernuzzi}}, \ and\ \bibinfo {author}
  {\bibfnamefont {S.}~\bibnamefont {Ansoldi}},\ }\href {\doibase
  10.1140/epja/s10050-022-00743-5} {\  (\bibinfo {year} {2021}),\
  10.1140/epja/s10050-022-00743-5},\ \Eprint {http://arxiv.org/abs/2111.13005}
  {arXiv:2111.13005 [astro-ph.HE]} \BibitemShut {NoStop}%
\bibitem [{\citenamefont {Radice}\ \emph {et~al.}(2022)\citenamefont {Radice},
  \citenamefont {Bernuzzi}, \citenamefont {Perego},\ and\ \citenamefont
  {Haas}}]{Radice:2021jtw}%
  \BibitemOpen
  \bibfield  {author} {\bibinfo {author} {\bibfnamefont {D.}~\bibnamefont
  {Radice}}, \bibinfo {author} {\bibfnamefont {S.}~\bibnamefont {Bernuzzi}},
  \bibinfo {author} {\bibfnamefont {A.}~\bibnamefont {Perego}}, \ and\ \bibinfo
  {author} {\bibfnamefont {R.}~\bibnamefont {Haas}},\ }\href {\doibase
  10.1093/mnras/stac589} {\bibfield  {journal} {\bibinfo  {journal} {Mon. Not.
  Roy. Astron. Soc.}\ }\textbf {\bibinfo {volume} {512}},\ \bibinfo {pages}
  {1499} (\bibinfo {year} {2022})},\ \Eprint {http://arxiv.org/abs/2111.14858}
  {arXiv:2111.14858 [astro-ph.HE]} \BibitemShut {NoStop}%
\bibitem [{\citenamefont {Wanajo}\ \emph {et~al.}(2014)\citenamefont {Wanajo},
  \citenamefont {Sekiguchi}, \citenamefont {Nishimura}, \citenamefont {Kiuchi},
  \citenamefont {Kyutoku},\ and\ \citenamefont {Shibata}}]{Wanajo:2014wha}%
  \BibitemOpen
  \bibfield  {author} {\bibinfo {author} {\bibfnamefont {S.}~\bibnamefont
  {Wanajo}}, \bibinfo {author} {\bibfnamefont {Y.}~\bibnamefont {Sekiguchi}},
  \bibinfo {author} {\bibfnamefont {N.}~\bibnamefont {Nishimura}}, \bibinfo
  {author} {\bibfnamefont {K.}~\bibnamefont {Kiuchi}}, \bibinfo {author}
  {\bibfnamefont {K.}~\bibnamefont {Kyutoku}}, \ and\ \bibinfo {author}
  {\bibfnamefont {M.}~\bibnamefont {Shibata}},\ }\href {\doibase
  10.1088/2041-8205/789/2/L39} {\bibfield  {journal} {\bibinfo  {journal}
  {Astrophys. J. Lett.}\ }\textbf {\bibinfo {volume} {789}},\ \bibinfo {pages}
  {L39} (\bibinfo {year} {2014})},\ \Eprint {http://arxiv.org/abs/1402.7317}
  {arXiv:1402.7317 [astro-ph.SR]} \BibitemShut {NoStop}%
\bibitem [{\citenamefont {Perego}\ \emph {et~al.}(2014)\citenamefont {Perego},
  \citenamefont {Rosswog}, \citenamefont {Cabez\'on}, \citenamefont {Korobkin},
  \citenamefont {K\"appeli}, \citenamefont {Arcones},\ and\ \citenamefont
  {Liebend\"orfer}}]{Perego:2014fma}%
  \BibitemOpen
  \bibfield  {author} {\bibinfo {author} {\bibfnamefont {A.}~\bibnamefont
  {Perego}}, \bibinfo {author} {\bibfnamefont {S.}~\bibnamefont {Rosswog}},
  \bibinfo {author} {\bibfnamefont {R.~M.}\ \bibnamefont {Cabez\'on}}, \bibinfo
  {author} {\bibfnamefont {O.}~\bibnamefont {Korobkin}}, \bibinfo {author}
  {\bibfnamefont {R.}~\bibnamefont {K\"appeli}}, \bibinfo {author}
  {\bibfnamefont {A.}~\bibnamefont {Arcones}}, \ and\ \bibinfo {author}
  {\bibfnamefont {M.}~\bibnamefont {Liebend\"orfer}},\ }\href {\doibase
  10.1093/mnras/stu1352} {\bibfield  {journal} {\bibinfo  {journal} {Mon. Not.
  Roy. Astron. Soc.}\ }\textbf {\bibinfo {volume} {443}},\ \bibinfo {pages}
  {3134} (\bibinfo {year} {2014})},\ \Eprint {http://arxiv.org/abs/1405.6730}
  {arXiv:1405.6730 [astro-ph.HE]} \BibitemShut {NoStop}%
\bibitem [{\citenamefont {Zappa}\ \emph {et~al.}(2022)\citenamefont {Zappa},
  \citenamefont {Bernuzzi}, \citenamefont {Radice},\ and\ \citenamefont
  {Perego}}]{Zappa:2022rpd}%
  \BibitemOpen
  \bibfield  {author} {\bibinfo {author} {\bibfnamefont {F.}~\bibnamefont
  {Zappa}}, \bibinfo {author} {\bibfnamefont {S.}~\bibnamefont {Bernuzzi}},
  \bibinfo {author} {\bibfnamefont {D.}~\bibnamefont {Radice}}, \ and\ \bibinfo
  {author} {\bibfnamefont {A.}~\bibnamefont {Perego}},\ }\href {\doibase
  10.1093/mnras/stad107} {\  (\bibinfo {year} {2022}),\
  10.1093/mnras/stad107},\ \Eprint {http://arxiv.org/abs/2210.11491}
  {arXiv:2210.11491 [astro-ph.HE]} \BibitemShut {NoStop}%
\bibitem [{\citenamefont {Metzger}\ and\ \citenamefont
  {Fern\'andez}(2014)}]{Metzger:2014ila}%
  \BibitemOpen
  \bibfield  {author} {\bibinfo {author} {\bibfnamefont {B.~D.}\ \bibnamefont
  {Metzger}}\ and\ \bibinfo {author} {\bibfnamefont {R.}~\bibnamefont
  {Fern\'andez}},\ }\href {\doibase 10.1093/mnras/stu802} {\bibfield  {journal}
  {\bibinfo  {journal} {Mon. Not. Roy. Astron. Soc.}\ }\textbf {\bibinfo
  {volume} {441}},\ \bibinfo {pages} {3444} (\bibinfo {year} {2014})},\ \Eprint
  {http://arxiv.org/abs/1402.4803} {arXiv:1402.4803 [astro-ph.HE]} \BibitemShut
  {NoStop}%
\bibitem [{\citenamefont {Martin}\ \emph {et~al.}(2015)\citenamefont {Martin},
  \citenamefont {Perego}, \citenamefont {Arcones}, \citenamefont {Thielemann},
  \citenamefont {Korobkin},\ and\ \citenamefont {Rosswog}}]{Martin:2015hxa}%
  \BibitemOpen
  \bibfield  {author} {\bibinfo {author} {\bibfnamefont {D.}~\bibnamefont
  {Martin}}, \bibinfo {author} {\bibfnamefont {A.}~\bibnamefont {Perego}},
  \bibinfo {author} {\bibfnamefont {A.}~\bibnamefont {Arcones}}, \bibinfo
  {author} {\bibfnamefont {F.-K.}\ \bibnamefont {Thielemann}}, \bibinfo
  {author} {\bibfnamefont {O.}~\bibnamefont {Korobkin}}, \ and\ \bibinfo
  {author} {\bibfnamefont {S.}~\bibnamefont {Rosswog}},\ }\href {\doibase
  10.1088/0004-637X/813/1/2} {\bibfield  {journal} {\bibinfo  {journal}
  {Astrophys. J.}\ }\textbf {\bibinfo {volume} {813}},\ \bibinfo {pages} {2}
  (\bibinfo {year} {2015})},\ \Eprint {http://arxiv.org/abs/1506.05048}
  {arXiv:1506.05048 [astro-ph.SR]} \BibitemShut {NoStop}%
\bibitem [{\citenamefont {Perego}\ \emph {et~al.}(2017)\citenamefont {Perego},
  \citenamefont {Radice},\ and\ \citenamefont {Bernuzzi}}]{Perego:2017wtu}%
  \BibitemOpen
  \bibfield  {author} {\bibinfo {author} {\bibfnamefont {A.}~\bibnamefont
  {Perego}}, \bibinfo {author} {\bibfnamefont {D.}~\bibnamefont {Radice}}, \
  and\ \bibinfo {author} {\bibfnamefont {S.}~\bibnamefont {Bernuzzi}},\ }\href
  {\doibase 10.3847/2041-8213/aa9ab9} {\bibfield  {journal} {\bibinfo
  {journal} {Astrophys. J. Lett.}\ }\textbf {\bibinfo {volume} {850}},\
  \bibinfo {pages} {L37} (\bibinfo {year} {2017})},\ \Eprint
  {http://arxiv.org/abs/1711.03982} {arXiv:1711.03982 [astro-ph.HE]}
  \BibitemShut {NoStop}%
\bibitem [{\citenamefont {Radice}\ \emph {et~al.}(2018)\citenamefont {Radice},
  \citenamefont {Perego}, \citenamefont {Hotokezaka}, \citenamefont {Fromm},
  \citenamefont {Bernuzzi},\ and\ \citenamefont {Roberts}}]{Radice:2018pdn}%
  \BibitemOpen
  \bibfield  {author} {\bibinfo {author} {\bibfnamefont {D.}~\bibnamefont
  {Radice}}, \bibinfo {author} {\bibfnamefont {A.}~\bibnamefont {Perego}},
  \bibinfo {author} {\bibfnamefont {K.}~\bibnamefont {Hotokezaka}}, \bibinfo
  {author} {\bibfnamefont {S.~A.}\ \bibnamefont {Fromm}}, \bibinfo {author}
  {\bibfnamefont {S.}~\bibnamefont {Bernuzzi}}, \ and\ \bibinfo {author}
  {\bibfnamefont {L.~F.}\ \bibnamefont {Roberts}},\ }\href {\doibase
  10.3847/1538-4357/aaf054} {\bibfield  {journal} {\bibinfo  {journal}
  {Astrophys. J.}\ }\textbf {\bibinfo {volume} {869}},\ \bibinfo {pages} {130}
  (\bibinfo {year} {2018})},\ \Eprint {http://arxiv.org/abs/1809.11161}
  {arXiv:1809.11161 [astro-ph.HE]} \BibitemShut {NoStop}%
\bibitem [{\citenamefont {Schmitt}\ and\ \citenamefont
  {Shternin}(2018)}]{Schmitt:2017efp}%
  \BibitemOpen
  \bibfield  {author} {\bibinfo {author} {\bibfnamefont {A.}~\bibnamefont
  {Schmitt}}\ and\ \bibinfo {author} {\bibfnamefont {P.}~\bibnamefont
  {Shternin}},\ }\href {\doibase 10.1007/978-3-319-97616-7_9} {\bibfield
  {journal} {\bibinfo  {journal} {Astrophys. Space Sci. Libr.}\ }\textbf
  {\bibinfo {volume} {457}},\ \bibinfo {pages} {455} (\bibinfo {year}
  {2018})},\ \Eprint {http://arxiv.org/abs/1711.06520} {arXiv:1711.06520
  [astro-ph.HE]} \BibitemShut {NoStop}%
\bibitem [{\citenamefont {Alford}\ \emph {et~al.}(2019)\citenamefont {Alford},
  \citenamefont {Harutyunyan},\ and\ \citenamefont
  {Sedrakian}}]{Alford:2019kdw}%
  \BibitemOpen
  \bibfield  {author} {\bibinfo {author} {\bibfnamefont {M.}~\bibnamefont
  {Alford}}, \bibinfo {author} {\bibfnamefont {A.}~\bibnamefont {Harutyunyan}},
  \ and\ \bibinfo {author} {\bibfnamefont {A.}~\bibnamefont {Sedrakian}},\
  }\href {\doibase 10.1103/PhysRevD.100.103021} {\bibfield  {journal} {\bibinfo
   {journal} {Phys. Rev. D}\ }\textbf {\bibinfo {volume} {100}},\ \bibinfo
  {pages} {103021} (\bibinfo {year} {2019})},\ \Eprint
  {http://arxiv.org/abs/1907.04192} {arXiv:1907.04192 [astro-ph.HE]}
  \BibitemShut {NoStop}%
\bibitem [{\citenamefont {Camelio}\ \emph
  {et~al.}(2023{\natexlab{a}})\citenamefont {Camelio}, \citenamefont
  {Gavassino}, \citenamefont {Antonelli}, \citenamefont {Bernuzzi},\ and\
  \citenamefont {Haskell}}]{Camelio:2022ljs}%
  \BibitemOpen
  \bibfield  {author} {\bibinfo {author} {\bibfnamefont {G.}~\bibnamefont
  {Camelio}}, \bibinfo {author} {\bibfnamefont {L.}~\bibnamefont {Gavassino}},
  \bibinfo {author} {\bibfnamefont {M.}~\bibnamefont {Antonelli}}, \bibinfo
  {author} {\bibfnamefont {S.}~\bibnamefont {Bernuzzi}}, \ and\ \bibinfo
  {author} {\bibfnamefont {B.}~\bibnamefont {Haskell}},\ }\href {\doibase
  10.1103/PhysRevD.107.103031} {\bibfield  {journal} {\bibinfo  {journal}
  {Phys. Rev. D}\ }\textbf {\bibinfo {volume} {107}},\ \bibinfo {pages}
  {103031} (\bibinfo {year} {2023}{\natexlab{a}})},\ \Eprint
  {http://arxiv.org/abs/2204.11809} {arXiv:2204.11809 [gr-qc]} \BibitemShut
  {NoStop}%
\bibitem [{\citenamefont {Camelio}\ \emph
  {et~al.}(2023{\natexlab{b}})\citenamefont {Camelio}, \citenamefont
  {Gavassino}, \citenamefont {Antonelli}, \citenamefont {Bernuzzi},\ and\
  \citenamefont {Haskell}}]{Camelio:2022fds}%
  \BibitemOpen
  \bibfield  {author} {\bibinfo {author} {\bibfnamefont {G.}~\bibnamefont
  {Camelio}}, \bibinfo {author} {\bibfnamefont {L.}~\bibnamefont {Gavassino}},
  \bibinfo {author} {\bibfnamefont {M.}~\bibnamefont {Antonelli}}, \bibinfo
  {author} {\bibfnamefont {S.}~\bibnamefont {Bernuzzi}}, \ and\ \bibinfo
  {author} {\bibfnamefont {B.}~\bibnamefont {Haskell}},\ }\href {\doibase
  10.1103/PhysRevD.107.103032} {\bibfield  {journal} {\bibinfo  {journal}
  {Phys. Rev. D}\ }\textbf {\bibinfo {volume} {107}},\ \bibinfo {pages}
  {103032} (\bibinfo {year} {2023}{\natexlab{b}})},\ \Eprint
  {http://arxiv.org/abs/2204.11810} {arXiv:2204.11810 [gr-qc]} \BibitemShut
  {NoStop}%
\bibitem [{\citenamefont {Alford}\ \emph {et~al.}(2018)\citenamefont {Alford},
  \citenamefont {Bovard}, \citenamefont {Hanauske}, \citenamefont {Rezzolla},\
  and\ \citenamefont {Schwenzer}}]{Alford:2017rxf}%
  \BibitemOpen
  \bibfield  {author} {\bibinfo {author} {\bibfnamefont {M.~G.}\ \bibnamefont
  {Alford}}, \bibinfo {author} {\bibfnamefont {L.}~\bibnamefont {Bovard}},
  \bibinfo {author} {\bibfnamefont {M.}~\bibnamefont {Hanauske}}, \bibinfo
  {author} {\bibfnamefont {L.}~\bibnamefont {Rezzolla}}, \ and\ \bibinfo
  {author} {\bibfnamefont {K.}~\bibnamefont {Schwenzer}},\ }\href {\doibase
  10.1103/PhysRevLett.120.041101} {\bibfield  {journal} {\bibinfo  {journal}
  {Phys. Rev. Lett.}\ }\textbf {\bibinfo {volume} {120}},\ \bibinfo {pages}
  {041101} (\bibinfo {year} {2018})},\ \Eprint
  {http://arxiv.org/abs/1707.09475} {arXiv:1707.09475 [gr-qc]} \BibitemShut
  {NoStop}%
\bibitem [{\citenamefont {Endrizzi}\ \emph {et~al.}(2020)\citenamefont
  {Endrizzi}, \citenamefont {Perego}, \citenamefont {Fabbri}, \citenamefont
  {Branca}, \citenamefont {Radice}, \citenamefont {Bernuzzi}, \citenamefont
  {Giacomazzo}, \citenamefont {Pederiva},\ and\ \citenamefont
  {Lovato}}]{Endrizzi:2019trv}%
  \BibitemOpen
  \bibfield  {author} {\bibinfo {author} {\bibfnamefont {A.}~\bibnamefont
  {Endrizzi}}, \bibinfo {author} {\bibfnamefont {A.}~\bibnamefont {Perego}},
  \bibinfo {author} {\bibfnamefont {F.~M.}\ \bibnamefont {Fabbri}}, \bibinfo
  {author} {\bibfnamefont {L.}~\bibnamefont {Branca}}, \bibinfo {author}
  {\bibfnamefont {D.}~\bibnamefont {Radice}}, \bibinfo {author} {\bibfnamefont
  {S.}~\bibnamefont {Bernuzzi}}, \bibinfo {author} {\bibfnamefont
  {B.}~\bibnamefont {Giacomazzo}}, \bibinfo {author} {\bibfnamefont
  {F.}~\bibnamefont {Pederiva}}, \ and\ \bibinfo {author} {\bibfnamefont
  {A.}~\bibnamefont {Lovato}},\ }\href {\doibase
  10.1140/epja/s10050-019-00018-6} {\bibfield  {journal} {\bibinfo  {journal}
  {Eur. Phys. J. A}\ }\textbf {\bibinfo {volume} {56}},\ \bibinfo {pages} {15}
  (\bibinfo {year} {2020})},\ \Eprint {http://arxiv.org/abs/1908.04952}
  {arXiv:1908.04952 [astro-ph.HE]} \BibitemShut {NoStop}%
\bibitem [{\citenamefont {Hammond}\ \emph {et~al.}(2021)\citenamefont
  {Hammond}, \citenamefont {Hawke},\ and\ \citenamefont
  {Andersson}}]{Hammond:2021vtv}%
  \BibitemOpen
  \bibfield  {author} {\bibinfo {author} {\bibfnamefont {P.}~\bibnamefont
  {Hammond}}, \bibinfo {author} {\bibfnamefont {I.}~\bibnamefont {Hawke}}, \
  and\ \bibinfo {author} {\bibfnamefont {N.}~\bibnamefont {Andersson}},\ }\href
  {\doibase 10.1103/PhysRevD.104.103006} {\bibfield  {journal} {\bibinfo
  {journal} {Phys. Rev. D}\ }\textbf {\bibinfo {volume} {104}},\ \bibinfo
  {pages} {103006} (\bibinfo {year} {2021})},\ \Eprint
  {http://arxiv.org/abs/2108.08649} {arXiv:2108.08649 [astro-ph.HE]}
  \BibitemShut {NoStop}%
\bibitem [{\citenamefont {Most}\ \emph {et~al.}(2022)\citenamefont {Most},
  \citenamefont {Haber}, \citenamefont {Harris}, \citenamefont {Zhang},
  \citenamefont {Alford},\ and\ \citenamefont {Noronha}}]{Most:2022yhe}%
  \BibitemOpen
  \bibfield  {author} {\bibinfo {author} {\bibfnamefont {E.~R.}\ \bibnamefont
  {Most}}, \bibinfo {author} {\bibfnamefont {A.}~\bibnamefont {Haber}},
  \bibinfo {author} {\bibfnamefont {S.~P.}\ \bibnamefont {Harris}}, \bibinfo
  {author} {\bibfnamefont {Z.}~\bibnamefont {Zhang}}, \bibinfo {author}
  {\bibfnamefont {M.~G.}\ \bibnamefont {Alford}}, \ and\ \bibinfo {author}
  {\bibfnamefont {J.}~\bibnamefont {Noronha}},\ }\href@noop {} {\  (\bibinfo
  {year} {2022})},\ \Eprint {http://arxiv.org/abs/2207.00442} {arXiv:2207.00442
  [astro-ph.HE]} \BibitemShut {NoStop}%
\bibitem [{\citenamefont {Longo~Micchi}\ \emph {et~al.}(2023)\citenamefont
  {Longo~Micchi}, \citenamefont {Radice},\ and\ \citenamefont
  {Chirenti}}]{LongoMicchi:2023khv}%
  \BibitemOpen
  \bibfield  {author} {\bibinfo {author} {\bibfnamefont {L.~F.}\ \bibnamefont
  {Longo~Micchi}}, \bibinfo {author} {\bibfnamefont {D.}~\bibnamefont
  {Radice}}, \ and\ \bibinfo {author} {\bibfnamefont {C.}~\bibnamefont
  {Chirenti}},\ }\href {\doibase 10.1093/mnras/stad2420} {\bibfield  {journal}
  {\bibinfo  {journal} {Mon. Not. Roy. Astron. Soc.}\ }\textbf {\bibinfo
  {volume} {525}},\ \bibinfo {pages} {6359} (\bibinfo {year} {2023})},\ \Eprint
  {http://arxiv.org/abs/2306.04711} {arXiv:2306.04711 [astro-ph.HE]}
  \BibitemShut {NoStop}%
\bibitem [{\citenamefont {Shibata}\ \emph {et~al.}(2011)\citenamefont
  {Shibata}, \citenamefont {Kiuchi}, \citenamefont {Sekiguchi},\ and\
  \citenamefont {Suwa}}]{Shibata:2011kx}%
  \BibitemOpen
  \bibfield  {author} {\bibinfo {author} {\bibfnamefont {M.}~\bibnamefont
  {Shibata}}, \bibinfo {author} {\bibfnamefont {K.}~\bibnamefont {Kiuchi}},
  \bibinfo {author} {\bibfnamefont {Y.-i.}\ \bibnamefont {Sekiguchi}}, \ and\
  \bibinfo {author} {\bibfnamefont {Y.}~\bibnamefont {Suwa}},\ }\href {\doibase
  10.1143/PTP.125.1255} {\bibfield  {journal} {\bibinfo  {journal} {Prog.
  Theor. Phys.}\ }\textbf {\bibinfo {volume} {125}},\ \bibinfo {pages} {1255}
  (\bibinfo {year} {2011})},\ \Eprint {http://arxiv.org/abs/1104.3937}
  {arXiv:1104.3937 [astro-ph.HE]} \BibitemShut {NoStop}%
\bibitem [{\citenamefont {Foucart}\ \emph {et~al.}(2015)\citenamefont
  {Foucart}, \citenamefont {O?Connor}, \citenamefont {Roberts}, \citenamefont
  {Duez}, \citenamefont {Haas}, \citenamefont {Kidder}, \citenamefont {Ott},
  \citenamefont {Pfeiffer}, \citenamefont {Scheel},\ and\ \citenamefont
  {Szilagyi}}]{Foucart_2015}%
  \BibitemOpen
  \bibfield  {author} {\bibinfo {author} {\bibfnamefont {F.}~\bibnamefont
  {Foucart}}, \bibinfo {author} {\bibfnamefont {E.}~\bibnamefont {O?Connor}},
  \bibinfo {author} {\bibfnamefont {L.}~\bibnamefont {Roberts}}, \bibinfo
  {author} {\bibfnamefont {M.~D.}\ \bibnamefont {Duez}}, \bibinfo {author}
  {\bibfnamefont {R.}~\bibnamefont {Haas}}, \bibinfo {author} {\bibfnamefont
  {L.~E.}\ \bibnamefont {Kidder}}, \bibinfo {author} {\bibfnamefont {C.~D.}\
  \bibnamefont {Ott}}, \bibinfo {author} {\bibfnamefont {H.~P.}\ \bibnamefont
  {Pfeiffer}}, \bibinfo {author} {\bibfnamefont {M.~A.}\ \bibnamefont
  {Scheel}}, \ and\ \bibinfo {author} {\bibfnamefont {B.}~\bibnamefont
  {Szilagyi}},\ }\href {\doibase 10.1103/physrevd.91.124021} {\bibfield
  {journal} {\bibinfo  {journal} {Physical Review D}\ }\textbf {\bibinfo
  {volume} {91}} (\bibinfo {year} {2015}),\
  10.1103/physrevd.91.124021}\BibitemShut {NoStop}%
\bibitem [{\citenamefont {Radice}\ \emph {et~al.}(2016)\citenamefont {Radice},
  \citenamefont {Galeazzi}, \citenamefont {Lippuner}, \citenamefont {Roberts},
  \citenamefont {Ott},\ and\ \citenamefont {Rezzolla}}]{Radice:2016dwd}%
  \BibitemOpen
  \bibfield  {author} {\bibinfo {author} {\bibfnamefont {D.}~\bibnamefont
  {Radice}}, \bibinfo {author} {\bibfnamefont {F.}~\bibnamefont {Galeazzi}},
  \bibinfo {author} {\bibfnamefont {J.}~\bibnamefont {Lippuner}}, \bibinfo
  {author} {\bibfnamefont {L.~F.}\ \bibnamefont {Roberts}}, \bibinfo {author}
  {\bibfnamefont {C.~D.}\ \bibnamefont {Ott}}, \ and\ \bibinfo {author}
  {\bibfnamefont {L.}~\bibnamefont {Rezzolla}},\ }\href {\doibase
  10.1093/mnras/stw1227} {\bibfield  {journal} {\bibinfo  {journal} {Mon. Not.
  Roy. Astron. Soc.}\ }\textbf {\bibinfo {volume} {460}},\ \bibinfo {pages}
  {3255} (\bibinfo {year} {2016})},\ \Eprint {http://arxiv.org/abs/1601.02426}
  {arXiv:1601.02426 [astro-ph.HE]} \BibitemShut {NoStop}%
\bibitem [{\citenamefont {Radice}\ and\ \citenamefont
  {Rezzolla}(2012{\natexlab{a}})}]{WHiskyTHC1}%
  \BibitemOpen
  \bibfield  {author} {\bibinfo {author} {\bibfnamefont {D.}~\bibnamefont
  {Radice}}\ and\ \bibinfo {author} {\bibfnamefont {L.}~\bibnamefont
  {Rezzolla}},\ }\href {\doibase 10.1051/0004-6361/201219735} {\bibfield
  {journal} {\bibinfo  {journal} {Astronomy and Astrophysics}\ }\textbf
  {\bibinfo {volume} {547}},\ \bibinfo {pages} {A26} (\bibinfo {year}
  {2012}{\natexlab{a}})}\BibitemShut {NoStop}%
\bibitem [{\citenamefont {Radice}\ \emph {et~al.}(2013)\citenamefont {Radice},
  \citenamefont {Rezzolla},\ and\ \citenamefont {Galeazzi}}]{WhiskyTHC2}%
  \BibitemOpen
  \bibfield  {author} {\bibinfo {author} {\bibfnamefont {D.}~\bibnamefont
  {Radice}}, \bibinfo {author} {\bibfnamefont {L.}~\bibnamefont {Rezzolla}}, \
  and\ \bibinfo {author} {\bibfnamefont {F.}~\bibnamefont {Galeazzi}},\ }\href
  {\doibase 10.1093/mnrasl/slt137} {\bibfield  {journal} {\bibinfo  {journal}
  {Monthly Notices of the Royal Astronomical Society: Letters}\ }\textbf
  {\bibinfo {volume} {437}},\ \bibinfo {pages} {L46} (\bibinfo {year}
  {2013})}\BibitemShut {NoStop}%
\bibitem [{\citenamefont {Foucart}\ \emph
  {et~al.}(2016{\natexlab{c}})\citenamefont {Foucart}, \citenamefont {Haas},
  \citenamefont {Duez}, \citenamefont {O'Connor}, \citenamefont {Ott},
  \citenamefont {Roberts}, \citenamefont {Kidder}, \citenamefont {Lippuner},
  \citenamefont {Pfeiffer},\ and\ \citenamefont {Scheel}}]{Foucart:2015gaa}%
  \BibitemOpen
  \bibfield  {author} {\bibinfo {author} {\bibfnamefont {F.}~\bibnamefont
  {Foucart}}, \bibinfo {author} {\bibfnamefont {R.}~\bibnamefont {Haas}},
  \bibinfo {author} {\bibfnamefont {M.~D.}\ \bibnamefont {Duez}}, \bibinfo
  {author} {\bibfnamefont {E.}~\bibnamefont {O'Connor}}, \bibinfo {author}
  {\bibfnamefont {C.~D.}\ \bibnamefont {Ott}}, \bibinfo {author} {\bibfnamefont
  {L.}~\bibnamefont {Roberts}}, \bibinfo {author} {\bibfnamefont {L.~E.}\
  \bibnamefont {Kidder}}, \bibinfo {author} {\bibfnamefont {J.}~\bibnamefont
  {Lippuner}}, \bibinfo {author} {\bibfnamefont {H.~P.}\ \bibnamefont
  {Pfeiffer}}, \ and\ \bibinfo {author} {\bibfnamefont {M.~A.}\ \bibnamefont
  {Scheel}},\ }\href {\doibase 10.1103/PhysRevD.93.044019} {\bibfield
  {journal} {\bibinfo  {journal} {Phys. Rev. D}\ }\textbf {\bibinfo {volume}
  {93}},\ \bibinfo {pages} {044019} (\bibinfo {year} {2016}{\natexlab{c}})},\
  \Eprint {http://arxiv.org/abs/1510.06398} {arXiv:1510.06398 [astro-ph.HE]}
  \BibitemShut {NoStop}%
\bibitem [{\citenamefont {Bombaci}\ and\ \citenamefont
  {Logoteta}(2018)}]{Bombaci:2018ksa}%
  \BibitemOpen
  \bibfield  {author} {\bibinfo {author} {\bibfnamefont {I.}~\bibnamefont
  {Bombaci}}\ and\ \bibinfo {author} {\bibfnamefont {D.}~\bibnamefont
  {Logoteta}},\ }\href {\doibase 10.1051/0004-6361/201731604} {\bibfield
  {journal} {\bibinfo  {journal} {Astron. Astrophys.}\ }\textbf {\bibinfo
  {volume} {609}},\ \bibinfo {pages} {A128} (\bibinfo {year} {2018})},\ \Eprint
  {http://arxiv.org/abs/1805.11846} {arXiv:1805.11846 [astro-ph.HE]}
  \BibitemShut {NoStop}%
\bibitem [{\citenamefont {{Hempel}}\ and\ \citenamefont
  {{Schaffner-Bielich}}(2010)}]{Hempel2010}%
  \BibitemOpen
  \bibfield  {author} {\bibinfo {author} {\bibfnamefont {M.}~\bibnamefont
  {{Hempel}}}\ and\ \bibinfo {author} {\bibfnamefont {J.}~\bibnamefont
  {{Schaffner-Bielich}}},\ }\href {\doibase 10.1016/j.nuclphysa.2010.02.010}
  {\bibfield  {journal} {\bibinfo  {journal} {\nphysa}\ }\textbf {\bibinfo
  {volume} {837}},\ \bibinfo {pages} {210} (\bibinfo {year} {2010})},\ \Eprint
  {http://arxiv.org/abs/0911.4073} {arXiv:0911.4073 [nucl-th]} \BibitemShut
  {NoStop}%
\bibitem [{\citenamefont {Steiner}\ \emph {et~al.}(2010)\citenamefont
  {Steiner}, \citenamefont {Lattimer},\ and\ \citenamefont
  {Brown}}]{Steiner_2010}%
  \BibitemOpen
  \bibfield  {author} {\bibinfo {author} {\bibfnamefont {A.~W.}\ \bibnamefont
  {Steiner}}, \bibinfo {author} {\bibfnamefont {J.~M.}\ \bibnamefont
  {Lattimer}}, \ and\ \bibinfo {author} {\bibfnamefont {E.~F.}\ \bibnamefont
  {Brown}},\ }\href {\doibase 10.1088/0004-637x/722/1/33} {\bibfield  {journal}
  {\bibinfo  {journal} {The Astrophysical Journal}\ }\textbf {\bibinfo {volume}
  {722}},\ \bibinfo {pages} {33} (\bibinfo {year} {2010})}\BibitemShut
  {NoStop}%
\bibitem [{\citenamefont {Chabanat}\ \emph {et~al.}(1998)\citenamefont
  {Chabanat}, \citenamefont {Bonche}, \citenamefont {Haensel}, \citenamefont
  {Meyer},\ and\ \citenamefont {Schaeffer}}]{Chabanat:1997un}%
  \BibitemOpen
  \bibfield  {author} {\bibinfo {author} {\bibfnamefont {E.}~\bibnamefont
  {Chabanat}}, \bibinfo {author} {\bibfnamefont {P.}~\bibnamefont {Bonche}},
  \bibinfo {author} {\bibfnamefont {P.}~\bibnamefont {Haensel}}, \bibinfo
  {author} {\bibfnamefont {J.}~\bibnamefont {Meyer}}, \ and\ \bibinfo {author}
  {\bibfnamefont {R.}~\bibnamefont {Schaeffer}},\ }\href {\doibase
  10.1016/S0375-9474(98)00180-8} {\bibfield  {journal} {\bibinfo  {journal}
  {Nucl. Phys. A}\ }\textbf {\bibinfo {volume} {635}},\ \bibinfo {pages} {231}
  (\bibinfo {year} {1998})},\ \bibinfo {note} {[Erratum: Nucl.Phys.A 643,
  441--441 (1998)]}\BibitemShut {NoStop}%
\bibitem [{\citenamefont {Schneider}\ \emph {et~al.}(2017)\citenamefont
  {Schneider}, \citenamefont {Roberts},\ and\ \citenamefont
  {Ott}}]{PhysRevC.96.065802}%
  \BibitemOpen
  \bibfield  {author} {\bibinfo {author} {\bibfnamefont {A.~S.}\ \bibnamefont
  {Schneider}}, \bibinfo {author} {\bibfnamefont {L.~F.}\ \bibnamefont
  {Roberts}}, \ and\ \bibinfo {author} {\bibfnamefont {C.~D.}\ \bibnamefont
  {Ott}},\ }\href {\doibase 10.1103/PhysRevC.96.065802} {\bibfield  {journal}
  {\bibinfo  {journal} {Phys. Rev. C}\ }\textbf {\bibinfo {volume} {96}},\
  \bibinfo {pages} {065802} (\bibinfo {year} {2017})}\BibitemShut {NoStop}%
\bibitem [{\citenamefont {Cromartie}\ \emph {et~al.}(2019)\citenamefont
  {Cromartie} \emph {et~al.}}]{Cromartie:2019kug}%
  \BibitemOpen
  \bibfield  {author} {\bibinfo {author} {\bibfnamefont {H.~T.}\ \bibnamefont
  {Cromartie}} \emph {et~al.} (\bibinfo {collaboration} {NANOGrav}),\ }\href
  {\doibase 10.1038/s41550-019-0880-2} {\bibfield  {journal} {\bibinfo
  {journal} {Nature Astron.}\ }\textbf {\bibinfo {volume} {4}},\ \bibinfo
  {pages} {72} (\bibinfo {year} {2019})},\ \Eprint
  {http://arxiv.org/abs/1904.06759} {arXiv:1904.06759 [astro-ph.HE]}
  \BibitemShut {NoStop}%
\bibitem [{\citenamefont {Fonseca}\ \emph {et~al.}(2021)\citenamefont {Fonseca}
  \emph {et~al.}}]{Fonseca:2021wxt}%
  \BibitemOpen
  \bibfield  {author} {\bibinfo {author} {\bibfnamefont {E.}~\bibnamefont
  {Fonseca}} \emph {et~al.},\ }\href {\doibase 10.3847/2041-8213/ac03b8}
  {\bibfield  {journal} {\bibinfo  {journal} {Astrophys. J. Lett.}\ }\textbf
  {\bibinfo {volume} {915}},\ \bibinfo {pages} {L12} (\bibinfo {year}
  {2021})},\ \Eprint {http://arxiv.org/abs/2104.00880} {arXiv:2104.00880
  [astro-ph.HE]} \BibitemShut {NoStop}%
\bibitem [{\citenamefont {Abbott}\ \emph {et~al.}(2018)\citenamefont {Abbott}
  \emph {et~al.}}]{Abbott:2018exr}%
  \BibitemOpen
  \bibfield  {author} {\bibinfo {author} {\bibfnamefont {B.~P.}\ \bibnamefont
  {Abbott}} \emph {et~al.} (\bibinfo {collaboration} {Virgo, LIGO
  Scientific}),\ }\href@noop {} {\  (\bibinfo {year} {2018})},\ \Eprint
  {http://arxiv.org/abs/1805.11581} {arXiv:1805.11581 [gr-qc]} \BibitemShut
  {NoStop}%
\bibitem [{\citenamefont {Miller}\ \emph {et~al.}(2021)\citenamefont {Miller}
  \emph {et~al.}}]{Miller:2021qha}%
  \BibitemOpen
  \bibfield  {author} {\bibinfo {author} {\bibfnamefont {M.~C.}\ \bibnamefont
  {Miller}} \emph {et~al.},\ }\href@noop {} {\  (\bibinfo {year} {2021})},\
  \Eprint {http://arxiv.org/abs/2105.06979} {arXiv:2105.06979 [astro-ph.HE]}
  \BibitemShut {NoStop}%
\bibitem [{\citenamefont {Riley}\ \emph {et~al.}(2021)\citenamefont {Riley}
  \emph {et~al.}}]{Riley:2021pdl}%
  \BibitemOpen
  \bibfield  {author} {\bibinfo {author} {\bibfnamefont {T.~E.}\ \bibnamefont
  {Riley}} \emph {et~al.},\ }\href@noop {} {\  (\bibinfo {year} {2021})},\
  \Eprint {http://arxiv.org/abs/2105.06980} {arXiv:2105.06980 [astro-ph.HE]}
  \BibitemShut {NoStop}%
\bibitem [{\citenamefont {Bernuzzi}\ and\ \citenamefont
  {Hilditch}(2010)}]{Bernuzzi:2009ex}%
  \BibitemOpen
  \bibfield  {author} {\bibinfo {author} {\bibfnamefont {S.}~\bibnamefont
  {Bernuzzi}}\ and\ \bibinfo {author} {\bibfnamefont {D.}~\bibnamefont
  {Hilditch}},\ }\href {\doibase 10.1103/PhysRevD.81.084003} {\bibfield
  {journal} {\bibinfo  {journal} {Phys. Rev. D}\ }\textbf {\bibinfo {volume}
  {81}},\ \bibinfo {pages} {084003} (\bibinfo {year} {2010})},\ \Eprint
  {http://arxiv.org/abs/0912.2920} {arXiv:0912.2920 [gr-qc]} \BibitemShut
  {NoStop}%
\bibitem [{\citenamefont {Hilditch}\ \emph {et~al.}(2013)\citenamefont
  {Hilditch}, \citenamefont {Bernuzzi}, \citenamefont {Thierfelder},
  \citenamefont {Cao}, \citenamefont {Tichy},\ and\ \citenamefont
  {Bruegmann}}]{Hilditch:2012fp}%
  \BibitemOpen
  \bibfield  {author} {\bibinfo {author} {\bibfnamefont {D.}~\bibnamefont
  {Hilditch}}, \bibinfo {author} {\bibfnamefont {S.}~\bibnamefont {Bernuzzi}},
  \bibinfo {author} {\bibfnamefont {M.}~\bibnamefont {Thierfelder}}, \bibinfo
  {author} {\bibfnamefont {Z.}~\bibnamefont {Cao}}, \bibinfo {author}
  {\bibfnamefont {W.}~\bibnamefont {Tichy}}, \ and\ \bibinfo {author}
  {\bibfnamefont {B.}~\bibnamefont {Bruegmann}},\ }\href {\doibase
  10.1103/PhysRevD.88.084057} {\bibfield  {journal} {\bibinfo  {journal} {Phys.
  Rev. D}\ }\textbf {\bibinfo {volume} {88}},\ \bibinfo {pages} {084057}
  (\bibinfo {year} {2013})},\ \Eprint {http://arxiv.org/abs/1212.2901}
  {arXiv:1212.2901 [gr-qc]} \BibitemShut {NoStop}%
\bibitem [{\citenamefont {Radice}\ and\ \citenamefont
  {Rezzolla}(2012{\natexlab{b}})}]{Radice:2012cu}%
  \BibitemOpen
  \bibfield  {author} {\bibinfo {author} {\bibfnamefont {D.}~\bibnamefont
  {Radice}}\ and\ \bibinfo {author} {\bibfnamefont {L.}~\bibnamefont
  {Rezzolla}},\ }\href {\doibase 10.1051/0004-6361/201219735} {\bibfield
  {journal} {\bibinfo  {journal} {Astron. Astrophys.}\ }\textbf {\bibinfo
  {volume} {547}},\ \bibinfo {pages} {A26} (\bibinfo {year}
  {2012}{\natexlab{b}})},\ \Eprint {http://arxiv.org/abs/1206.6502}
  {arXiv:1206.6502 [astro-ph.IM]} \BibitemShut {NoStop}%
\bibitem [{\citenamefont {Radice}\ \emph
  {et~al.}(2014{\natexlab{a}})\citenamefont {Radice}, \citenamefont
  {Rezzolla},\ and\ \citenamefont {Galeazzi}}]{Radice:2013hxh}%
  \BibitemOpen
  \bibfield  {author} {\bibinfo {author} {\bibfnamefont {D.}~\bibnamefont
  {Radice}}, \bibinfo {author} {\bibfnamefont {L.}~\bibnamefont {Rezzolla}}, \
  and\ \bibinfo {author} {\bibfnamefont {F.}~\bibnamefont {Galeazzi}},\ }\href
  {\doibase 10.1093/mnrasl/slt137} {\bibfield  {journal} {\bibinfo  {journal}
  {Mon. Not. Roy. Astron. Soc.}\ }\textbf {\bibinfo {volume} {437}},\ \bibinfo
  {pages} {L46} (\bibinfo {year} {2014}{\natexlab{a}})},\ \Eprint
  {http://arxiv.org/abs/1306.6052} {arXiv:1306.6052 [gr-qc]} \BibitemShut
  {NoStop}%
\bibitem [{\citenamefont {Radice}\ \emph
  {et~al.}(2014{\natexlab{b}})\citenamefont {Radice}, \citenamefont
  {Rezzolla},\ and\ \citenamefont {Galeazzi}}]{Radice:2013xpa}%
  \BibitemOpen
  \bibfield  {author} {\bibinfo {author} {\bibfnamefont {D.}~\bibnamefont
  {Radice}}, \bibinfo {author} {\bibfnamefont {L.}~\bibnamefont {Rezzolla}}, \
  and\ \bibinfo {author} {\bibfnamefont {F.}~\bibnamefont {Galeazzi}},\ }\href
  {\doibase 10.1088/0264-9381/31/7/075012} {\bibfield  {journal} {\bibinfo
  {journal} {Class. Quant. Grav.}\ }\textbf {\bibinfo {volume} {31}},\ \bibinfo
  {pages} {075012} (\bibinfo {year} {2014}{\natexlab{b}})},\ \Eprint
  {http://arxiv.org/abs/1312.5004} {arXiv:1312.5004 [gr-qc]} \BibitemShut
  {NoStop}%
\bibitem [{\citenamefont {Radice}\ \emph {et~al.}(2015)\citenamefont {Radice},
  \citenamefont {Rezzolla},\ and\ \citenamefont {Galeazzi}}]{Radice:2015nva}%
  \BibitemOpen
  \bibfield  {author} {\bibinfo {author} {\bibfnamefont {D.}~\bibnamefont
  {Radice}}, \bibinfo {author} {\bibfnamefont {L.}~\bibnamefont {Rezzolla}}, \
  and\ \bibinfo {author} {\bibfnamefont {F.}~\bibnamefont {Galeazzi}},\
  }\href@noop {} {\bibfield  {journal} {\bibinfo  {journal} {ASP Conf. Ser.}\
  }\textbf {\bibinfo {volume} {498}},\ \bibinfo {pages} {121} (\bibinfo {year}
  {2015})},\ \Eprint {http://arxiv.org/abs/1502.00551} {arXiv:1502.00551
  [gr-qc]} \BibitemShut {NoStop}%
\bibitem [{\citenamefont {Font}\ \emph {et~al.}(2002)\citenamefont {Font},
  \citenamefont {Goodale}, \citenamefont {Iyer}, \citenamefont {Miller},
  \citenamefont {Rezzolla}, \citenamefont {Seidel}, \citenamefont
  {Stergioulas}, \citenamefont {Suen},\ and\ \citenamefont
  {Tobias}}]{Font:2001ew}%
  \BibitemOpen
  \bibfield  {author} {\bibinfo {author} {\bibfnamefont {J.~A.}\ \bibnamefont
  {Font}}, \bibinfo {author} {\bibfnamefont {T.}~\bibnamefont {Goodale}},
  \bibinfo {author} {\bibfnamefont {S.}~\bibnamefont {Iyer}}, \bibinfo {author}
  {\bibfnamefont {M.~A.}\ \bibnamefont {Miller}}, \bibinfo {author}
  {\bibfnamefont {L.}~\bibnamefont {Rezzolla}}, \bibinfo {author}
  {\bibfnamefont {E.}~\bibnamefont {Seidel}}, \bibinfo {author} {\bibfnamefont
  {N.}~\bibnamefont {Stergioulas}}, \bibinfo {author} {\bibfnamefont {W.-M.}\
  \bibnamefont {Suen}}, \ and\ \bibinfo {author} {\bibfnamefont
  {M.}~\bibnamefont {Tobias}},\ }\href {\doibase 10.1103/PhysRevD.65.084024}
  {\bibfield  {journal} {\bibinfo  {journal} {Phys. Rev. D}\ }\textbf {\bibinfo
  {volume} {65}},\ \bibinfo {pages} {084024} (\bibinfo {year} {2002})},\
  \Eprint {http://arxiv.org/abs/gr-qc/0110047} {arXiv:gr-qc/0110047}
  \BibitemShut {NoStop}%
\bibitem [{\citenamefont {Schnetter}\ \emph {et~al.}(2007)\citenamefont
  {Schnetter}, \citenamefont {Ott}, \citenamefont {Allen}, \citenamefont
  {Diener}, \citenamefont {Goodale}, \citenamefont {Radke}, \citenamefont
  {Seidel},\ and\ \citenamefont {Shalf}}]{Schnetter:2007rb}%
  \BibitemOpen
  \bibfield  {author} {\bibinfo {author} {\bibfnamefont {E.}~\bibnamefont
  {Schnetter}}, \bibinfo {author} {\bibfnamefont {C.~D.}\ \bibnamefont {Ott}},
  \bibinfo {author} {\bibfnamefont {G.}~\bibnamefont {Allen}}, \bibinfo
  {author} {\bibfnamefont {P.}~\bibnamefont {Diener}}, \bibinfo {author}
  {\bibfnamefont {T.}~\bibnamefont {Goodale}}, \bibinfo {author} {\bibfnamefont
  {T.}~\bibnamefont {Radke}}, \bibinfo {author} {\bibfnamefont
  {E.}~\bibnamefont {Seidel}}, \ and\ \bibinfo {author} {\bibfnamefont
  {J.}~\bibnamefont {Shalf}},\ }\href@noop {} {\  (\bibinfo {year} {2007})},\
  \Eprint {http://arxiv.org/abs/0707.1607} {arXiv:0707.1607 [cs.DC]}
  \BibitemShut {NoStop}%
\bibitem [{\citenamefont {Perego}\ \emph {et~al.}(2019)\citenamefont {Perego},
  \citenamefont {Bernuzzi},\ and\ \citenamefont {Radice}}]{Perego:2019adq}%
  \BibitemOpen
  \bibfield  {author} {\bibinfo {author} {\bibfnamefont {A.}~\bibnamefont
  {Perego}}, \bibinfo {author} {\bibfnamefont {S.}~\bibnamefont {Bernuzzi}}, \
  and\ \bibinfo {author} {\bibfnamefont {D.}~\bibnamefont {Radice}},\
  }\href@noop {} {\bibfield  {journal} {\bibinfo  {journal} {arXiv e-prints}\ }
  (\bibinfo {year} {2019})},\ \Eprint {http://arxiv.org/abs/1903.07898}
  {arXiv:1903.07898 [gr-qc]} \BibitemShut {NoStop}%
\bibitem [{\citenamefont {Loffredo}\ \emph {et~al.}(2022)\citenamefont
  {Loffredo}, \citenamefont {Perego}, \citenamefont {Logoteta},\ and\
  \citenamefont {Branchesi}}]{Loffredo:2022prq}%
  \BibitemOpen
  \bibfield  {author} {\bibinfo {author} {\bibfnamefont {E.}~\bibnamefont
  {Loffredo}}, \bibinfo {author} {\bibfnamefont {A.}~\bibnamefont {Perego}},
  \bibinfo {author} {\bibfnamefont {D.}~\bibnamefont {Logoteta}}, \ and\
  \bibinfo {author} {\bibfnamefont {M.}~\bibnamefont {Branchesi}},\ }\href@noop
  {} {\  (\bibinfo {year} {2022})},\ \Eprint {http://arxiv.org/abs/2209.04458}
  {arXiv:2209.04458 [astro-ph.HE]} \BibitemShut {NoStop}%
\bibitem [{\citenamefont {Alford}\ and\ \citenamefont
  {Haber}(2021)}]{Alford:2020pld}%
  \BibitemOpen
  \bibfield  {author} {\bibinfo {author} {\bibfnamefont {M.~G.}\ \bibnamefont
  {Alford}}\ and\ \bibinfo {author} {\bibfnamefont {A.}~\bibnamefont {Haber}},\
  }\href {\doibase 10.1103/PhysRevC.103.045810} {\bibfield  {journal} {\bibinfo
   {journal} {Phys. Rev. C}\ }\textbf {\bibinfo {volume} {103}},\ \bibinfo
  {pages} {045810} (\bibinfo {year} {2021})},\ \Eprint
  {http://arxiv.org/abs/2009.05181} {arXiv:2009.05181 [nucl-th]} \BibitemShut
  {NoStop}%
\bibitem [{\citenamefont {Alford}\ and\ \citenamefont
  {Schwenzer}(2014)}]{Alford:2013pma}%
  \BibitemOpen
  \bibfield  {author} {\bibinfo {author} {\bibfnamefont {M.~G.}\ \bibnamefont
  {Alford}}\ and\ \bibinfo {author} {\bibfnamefont {K.}~\bibnamefont
  {Schwenzer}},\ }\href {\doibase 10.1103/PhysRevLett.113.251102} {\bibfield
  {journal} {\bibinfo  {journal} {Phys. Rev. Lett.}\ }\textbf {\bibinfo
  {volume} {113}},\ \bibinfo {pages} {251102} (\bibinfo {year} {2014})},\
  \Eprint {http://arxiv.org/abs/1310.3524} {arXiv:1310.3524 [astro-ph.HE]}
  \BibitemShut {NoStop}%
\bibitem [{\citenamefont {{Alford}}\ \emph {et~al.}(2013)\citenamefont
  {{Alford}}, \citenamefont {{Han}},\ and\ \citenamefont
  {{Prakash}}}]{AlfordCSS13}%
  \BibitemOpen
  \bibfield  {author} {\bibinfo {author} {\bibfnamefont {M.~G.}\ \bibnamefont
  {{Alford}}}, \bibinfo {author} {\bibfnamefont {S.}~\bibnamefont {{Han}}}, \
  and\ \bibinfo {author} {\bibfnamefont {M.}~\bibnamefont {{Prakash}}},\ }\href
  {\doibase 10.1103/PhysRevD.88.083013} {\bibfield  {journal} {\bibinfo
  {journal} {\prd}\ }\textbf {\bibinfo {volume} {88}},\ \bibinfo {eid} {083013}
  (\bibinfo {year} {2013})}\BibitemShut {NoStop}%
\end{thebibliography}%

\clearpage
\appendix
\section{Supplemental Material}
\subsection{Details on the calculation of several post-process diagnostics}
\label{app:diagnostics}
As the EOS models which we employ in our study do not account for the presence of 
trapped neutrinos, we use the conditions extracted from our numerical simulations to 
infer the neutrino chemical potential used to calculate $\mu_\Delta^{npe\nu}$ (see 
Fig.~\ref{fig:betaequil_contours}). To do so, we take the following steps:
\begin{enumerate}
\item We calculate the Fermi-Dirac integral of order 2, $F_2$, using two 
forms. The first form assumes weak equilibrium and relates $F_2$ to the neutrino fraction analytically, 
following~\cite{Perego:2019adq},
\begin{equation}\label{eq:F2_ynu}
F_2(Y_\nu) = \dfrac{\rho (hc)^3}{4\pi m_{\rm b} (k_{\rm B} T)^3} Y_\nu,
\end{equation}
where $m_{\rm b}$ is the baryon mass.
Note that Eq.~\eqref{eq:F2_ynu} 
differs from the form used in~\cite{Perego:2019adq} by an exponential factor which 
suppresses the impact of neutrinos in low-density regions; we find that this 
exponential factor does not impact our results, so we opt to remove it for 
simplicity. 
\item We consider another form of $F_2$, which considers its standard 
definition in terms of the neutrino degeneracy 
parameter $\eta_\nu$,
\begin{equation}\label{eq:F2_eta}
F_2(\eta) = \int_0^\infty \dfrac{\varepsilon^2 {\rm d}\varepsilon}{1 + e^{(\varepsilon - \eta_\nu)}}\,.
\end{equation}
\item We evaluate Eq.~\eqref{eq:F2_eta} and interpolate using a finely sampled 
cubic spline. We then numerically invert the interpolated function to obtain 
$\eta_\nu(F_2)$.
\item We combine Eq.~\eqref{eq:F2_ynu} and Eq.~\eqref{eq:F2_eta} to obtain 
$\eta_\nu(Y_\nu) = \eta_\nu[F_2(Y_\nu)]$ numerically.
\item We calculate the neutrino chemical potential as
\begin{equation}\label{eq:eta_nu}
\mu_\nu = \eta_\nu(Y_\nu) T.
\end{equation}
\end{enumerate}
The use of Eqs.~\eqref{eq:F2_ynu}-\eqref{eq:eta_nu} to calculate $\mu_\nu$ 
is most accurate in 
regions with significant neutrino fractions $Y_\nu \gtrsim 10^{-4}$, which we find 
generally coincide with densities $\rho \gtrsim \SI{e12}{\g\per\cm\cubed}$. 

In Fig.~\ref{fig:pressure_eq} we show the deviation of the simulation pressure from 
that under the assumption of local weak equilibrium. The calculation of $P_{\rm eq}$ 
requires knowledge of the local temperature $T^*$ and electron fraction 
$Y_{\rm e}^*$ under the assumption of weak equilibrium, which in turn
requires calculation of the contribution of neutrinos to the total system energy 
density
\begin{equation}\label{eq:u_nu}
u_\nu = \dfrac{4\pi}{(h_{\rm P}c)^3}(k_{\rm B} T)^4\left[ \dfrac{7\pi^4}{20} + 
\dfrac{\eta_{\nu_e}^2}{2} \left(\pi^2 + \eta_{\nu_e}^2  \right)  \right],
\end{equation}
and the difference of neutrino and anti-neutrino fractions
\begin{equation}\label{eq:Y_nu}
Y_{\nu_e} - Y_{\bar{\nu}_e} = \dfrac{4\pi m_{\rm b}}{3\rho (h_{\rm P}c)^3}(k_{\rm B} T)^3 \eta_{\nu_e} (\pi^2 + \eta_{\nu_e}),
\end{equation}
where $n_{\rm b}$ is the baryon number density and
$\eta_{\nu_e} = \eta_{\nu_e} (\rho, T^*, Y_{\rm e}^*)$.
In order to
calculate $T^*$ and $Y_{\rm e}^*$ we numerically invert the relationships established
by Eqs.~\eqref{eq:u_nu}-\eqref{eq:Y_nu} (along with the definitions established in 
Eqs.~\eqref{eq:u_total_def}-\eqref{eq:Ylep_def}) 
using Newton-Raphson root-finding. We
then calculate $P_{\rm eq}(\rho, T^*, Y_{\rm e}^*)$ by interpolating the EOS model
pressure to the inferred temperature and electron fraction.

\subsection{Post-process analyses with different EOS, mass, ratio, and grid resolution}
\label{app:EOS_res_q}
\begin{figure}[b]
\includegraphics[scale=1.2]{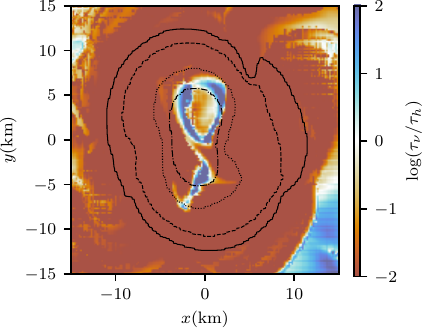}
\centering
\caption{
Equatorial snapshot of the ratio of neutrino 
interaction timescale to density oscillation timescale for the unequal mass ratio 
($q=1.2$), LR simulation with use of the DD2 EOS. The neutrino interaction timescales 
are much faster than the 
timescale of local density oscillations, consistent with result for the equal mass, 
SR simulation with use of the same EOS (see Fig.~\ref{fig:pressure_eq}).
}
\label{fig:timescales_pav}
\end{figure}

\begin{figure*}[htb]
\includegraphics{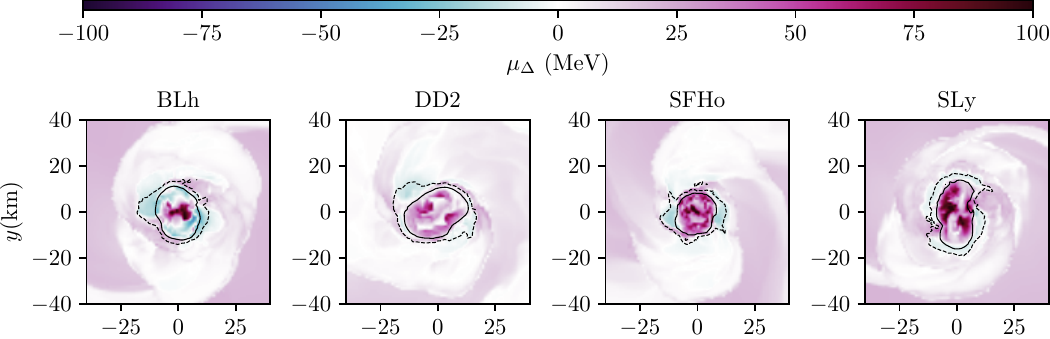}\\
\includegraphics{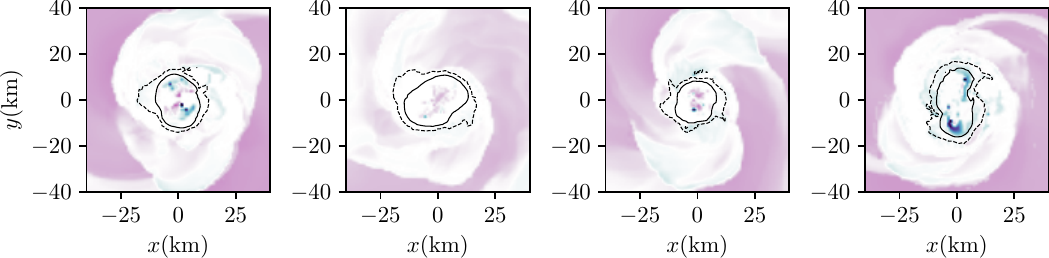}\\
\includegraphics[scale=0.31]{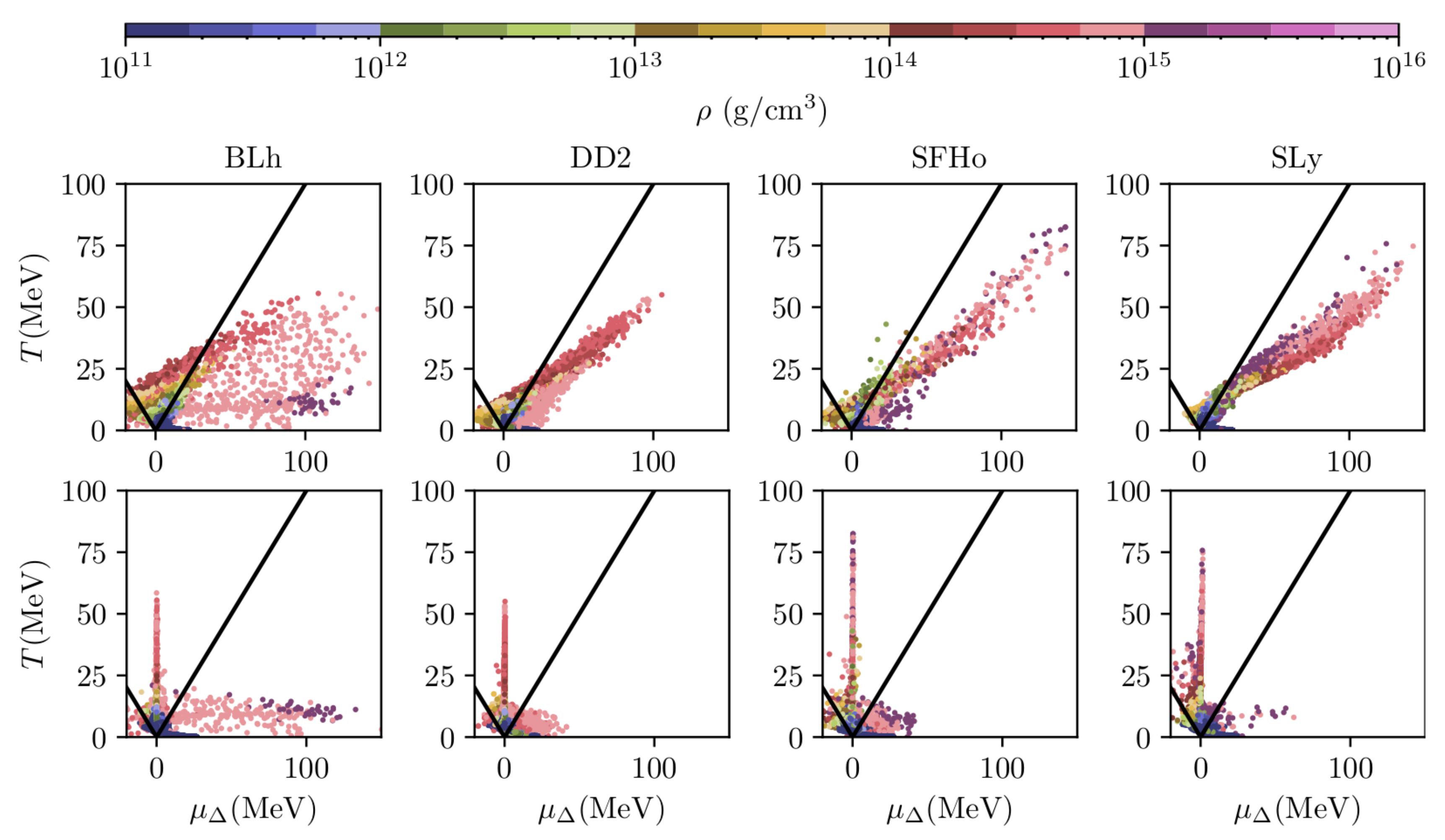}
\centering
\caption{
{\it Top panels:} Equatorial snapshots of the out-of-equilibrium chemical potential 
$\mu_\Delta$ at a time shortly after merger for the equal mass, SR simulations 
in our work. From left to right, we depict results for simulations which 
use the BLh, DD2, SFHo, and SLy EOSs. The top panel depicts the 
approximate chemical potential assuming \emph{neutrinoless} $\beta$-decay
($\mu_\Delta^{npe} = \mu_{\rm n} - \mu_{\rm p} - \mu_{\rm e}$). The lower 
panel includes the contribution from neutrinos 
($\mu_\Delta^{npe\nu} = \mu_{\rm n} - \mu_{\rm p} - \mu_{\rm e} - 
\mu_{\bar{\nu}}$). We highlight regions in 
the post-merger environment with $\rho \geq \SI{e13}{g\per\cm\cubed}$ and 
$\rho \geq \SI{e14}{g\per\cm\cubed}$ using dashed and solid black lines, 
respectively. 
{\it Bottom panels:} Phase-space ($\mu_\Delta$-$T$ plane) histograms for the same 
simulations depicted in the top panels
The colorbar corresponds to the rest mass density. For each
simulation we depict the state of matter over a fixed time window of approximately 
3 ms before and after the merger. The $\mu_\Delta = T$ condition is shown using a 
solid black line.
}
\label{fig:betaequil_contours_EOS}
\end{figure*}

\begin{figure*}[htb]
\includegraphics[scale=0.8]{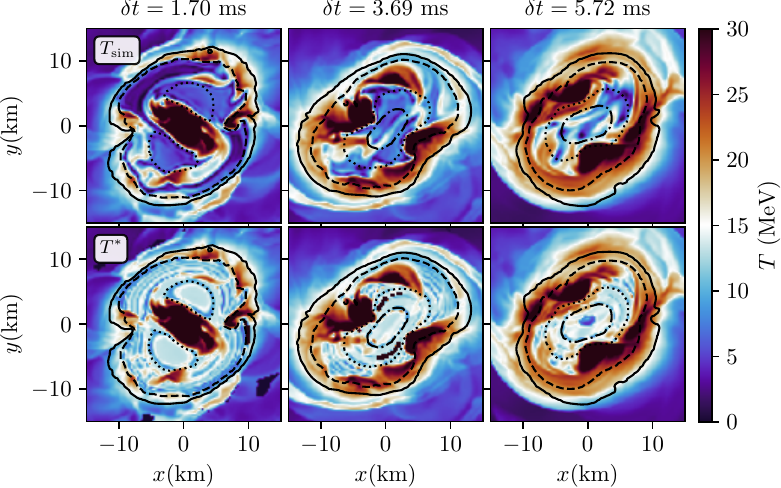}
\includegraphics[scale=0.78]{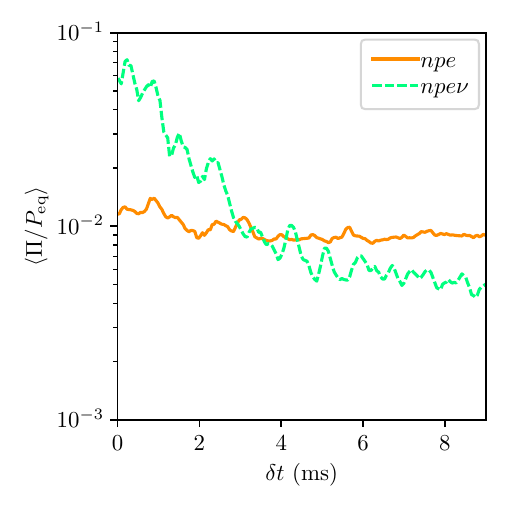}
\centering
\caption{
{\it Left:} Equatorial snapshots of the simulation pressure $T_{\rm sim}$ (top 
panels) and the pressure inferred when assuming a trapped neutrino gas $T^*$ (bottom 
panels) for the equal mass ratio, SR simulation 
with use of the DD2 EOS (the same simulation depicted in 
Fig.~\ref{fig:betaequil_contours}). 
{\it Right:} Rest mass density weighted spatial average of the pressure deviation 
from equilibrium under the assumption of $npe$ (solid orange line) and $npe\nu$ 
(dashed green line) matter, for the same simulation depicted in the left panel.
}
\label{fig:T_Tstar}
\end{figure*}
In the main text we showcase results for the equal mass ratio, SR simulation with 
use of the DD2 EOS. 
In Fig.~\ref{fig:timescales_pav} we 
show the ratio of neutrino interaction timescale to density oscillation 
timescale for the unequal mass ratio, LR simulation with use of the DD2 EOS. We find 
that, regardless of mass ratio or grid resolution considered, 
neutrino interactions occur much 
faster than the local density oscillations. We expect that neutrino interactions 
occur on the weak interaction timescale 
$10^{-10} - 10^{-8}$ s~\cite{Hammond:2021vtv}, 
which is faster than what is resolved in our 
simulations ($\Delta t \approx \SI{8.22e-8}{s}$ and 
$\Delta t \approx \SI{6.16e-8}{s}$ for the LR and SR simulations, respectively) and 
as such we expect the matter to remain close to local weak equilibrium at these 
resolutions.
In Fig.~\ref{fig:betaequil_contours_EOS} we show 
several diagnostics which reproduce the results of the main text while varying the 
EOS. We find that the general results summarized 
in the {\it Conclusion} section of the main text hold regardless of EOS.
For instance, we find out-of-equilibrium chemical potentials which remain within 
$\mu_\Delta \lesssim \SI{2}{MeV}$ for the majority of the RMNS if we consider the 
post-merger system to consist in part of a trapped neutrino gas. There are 
high-density regions in the RMNS which show significant deviations from 
weak equilibrium. However, for all cases we consider, these regions tend to be small 
and transient, existing for only a few ms after the merger. We note that not all EOSs 
we consider lead to long-lived RMNSs for which out-of-equilibrium effects would be 
most relevant. Nevertheless, we find that the matter is driven toward weak 
equilibrium on sufficiently fast timescales for out-of-equilibrium effects to 
potentially be mitigated, regardless of RMNS lifetime. 

\subsection{Additional post-process analyses}
For additional analyses to support those presented in the main text, we consider 
qualitative diagnostics for the equal mass ratio, SR simulation with use of the DD2 
EOS. In Fig.~\ref{fig:T_Tstar} we show the temperature 
extracted from our simulations $T_{\rm sim}$ along with the inferred temperature 
$T^*$ assuming a neutrino trapped gas, using 
Eqs.~\eqref{eq:u_total_def}-\eqref{eq:Ylep_def} and 
Eqs.~\eqref{eq:u_nu}-\eqref{eq:Y_nu}. When we account for a neutrino trapped gas we 
note a relative increase in the local temperature in the core of the RMNS. The 
regions with $T^\ast > T_{\rm sim}$ spatially coincide with the translucent 
regions discussed in the main text and are where we see 
significant deviations from weak equilibrium as suggested by other diagnostics. The 
hot regions originating from the shocked interface at the merger are expected to 
copiously produce neutrinos~\cite{Zappa:2022rpd} which 
inundate the translucent regions. 
However, the neutrinos produced in hot regions 
cannot equilibrate due to Pauli blocking which suppresses heat exchange with the 
neutrino fluid~\cite{Perego:2019adq}.
In Fig.~\ref{fig:T_Tstar} we also show the $\rho$-weighted 
spatial average of the pressure 
deviation from equilibrium as a function of time for the equal 
mass ratio, SR simulation with use of the DD2 EOS. Fig.~\ref{fig:T_Tstar} 
shows that in the few ms immediately following the merger, the average deviation from 
equilibrium is significantly higher if we treat the equilibrium state to be $npe\nu$ 
matter, mostly due to the fact that the initial conditions for our simulations are 
for $npe$ matter in equilibrium, consistent with cold neutron stars during the 
inspiral. 
We note that the RMNS cores remain cold and close to the initial
conditions. 
As such, the state of the majority of the matter 
during the inspiral, at the merger, and immediately after the merger is expected to 
be closer to neutrino-less $\beta$-equilibrium. 
However, within $\sim \SI{4}{ms}$ after 
the merger, we find that the fluid pressure deviates significantly less from $npe\nu$ 
equilibrium than from $npe$ equilibrium, consistent with the picture 
of significant neutrino trapping.

\end{document}